%% file: main.tex
\pdfoutput=1

\documentclass[conference]{IEEEtran}
\IEEEoverridecommandlockouts

\usepackage{cite}
\usepackage{amsmath,amssymb,amsfonts}
\usepackage{graphicx}
\usepackage{caption}
\usepackage{textcomp}
\usepackage{xcolor}
\usepackage{ltl}
\usepackage{tikz, tikz-3dplot}
\usetikzlibrary{arrows,backgrounds,decorations,decorations.pathmorphing,
positioning,fit,automata,shapes,snakes,patterns,plotmarks,calc,trees,arrows.meta}
\usepackage{algorithm}
\usepackage{algorithmicx}
\usepackage{algpseudocode}
\usepackage{multicol}
\usepackage{wrapfig}
\usepackage{subcaption}
\usepackage{xspace}
\usepackage{pgfplots}
\pgfplotsset{compat=newest}
\usepackage{pgfplotstable}
\usepackage{pifont}
\usepackage{todonotes}
\usepackage{hyperref}
\hypersetup{%
	colorlinks=true,           
	allcolors=blue!70!black,   
	pdfstartview=Fit,          
	breaklinks=true,
	pdfauthor={gxjlsbh},
	pdftitle={Distributed Runtime Verification of Metric Temporal Properties}}

\def\BibTeX{{\rm B\kern-.05em{\sc i\kern-.025em b}\kern-.08em
    T\kern-.1667em\lower.7ex\hbox{E}\kern-.125emX}}
\begin{document}

\renewcommand{\baselinestretch}{.98}
\title{Distributed Runtime Verification of Metric Temporal Properties for Cross-Chain Protocols\\
}

\author{
	\IEEEauthorblockN{Ritam Ganguly\IEEEauthorrefmark{1}, Yingjie Xue\IEEEauthorrefmark{2}, Aaron Jonckheere\IEEEauthorrefmark{1}, Parker Ljung\IEEEauthorrefmark{2}, Benjamin Schornstein\IEEEauthorrefmark{2}, \\ Borzoo Bonakdarpour\IEEEauthorrefmark{1}, and Maurice Herlihy\IEEEauthorrefmark{2}}
	\IEEEauthorblockA{\IEEEauthorrefmark{1}Michigan State University
		\textsf{\{gangulyr, jonckh16, borzoo\}@msu.edu}}
	\IEEEauthorblockA{\IEEEauthorrefmark{2}Brown University
		\textsf{\{yingjie\_xue, parker\_ljung, benjamin\_schornstein, mph\}@brown.edu}}
}

%
%

\maketitle

\thispagestyle{plain}
\pagestyle{plain}

\newcommand{\MTL}{\textsf{\small MTL}\xspace}
\newcommand{\LTL}{\texttt{LTL}\xspace}
\newcommand{\FLTL}{\texttt{FLTL}\xspace}

\newcommand{\intervalSet}{\mathbb{I}}
\newcommand{\naturalSet}{\mathbb{N}}
\newcommand{\realSet}{\mathbb{R}}
\newcommand{\realPlusSet}{\mathbb{R}_{\geq 0}}
\newcommand{\wholeSet}{\mathbb{Z}}
\newcommand{\wholePlusSet}{\mathbb{Z}_{\geq 0}}

\newcommand{\qed}{$~\blacksquare$}

\newcommand{\interval}{\mathcal{I}}
\newcommand{\Pred}{\mathsf{AP}}
\newcommand{\p}{\mathsf{p}}
\newcommand{\Time}{\tau}
\newcommand{\StrucV}{\bar{\mathcal{D}}}
\newcommand{\TimeV}{\bar{\tau}}
\newcommand{\Istart}{\mathit{start}}
\newcommand{\Iend}{\mathit{end}}

\newcommand{\tru}{\mathtt{true}}
\newcommand{\fals}{\mathtt{false}}

\DeclareRobustCommand{\F}{\LTLdiamond}
\DeclareRobustCommand{\G}{\LTLsquare}
\DeclareRobustCommand{\U}{\,\mathcal U \,}
\DeclareRobustCommand{\X}{\LTLcircle}

\newcommand{\Proc}{\mathcal{P}}
\newcommand{\globalC}{\mathcal{G}}
\newcommand{\Events}{\mathcal{E}}
\newcommand{\hb}{\rightsquigarrow}
\newcommand{\cc}{\mathcal{C}}
\newcommand{\ccAll}{\mathbb{C}}
\newcommand{\front}{\mathsf{front}}
\newcommand{\Tr}{\mathsf{Tr}}
\newcommand{\seg}{\textit{seg}}
\newcommand{\trace}{\alpha}
\newcommand{\RTime}{\sigma}
\newcommand{\RTimeV}{\bar{\sigma}}
\newcommand{\LTime}{\pi}
\newcommand{\LTimeV}{\bar{\pi}}

\newcommand{\Pro}{\mathsf{Pr}}
\newcommand{\hbSet}{\mathsf{hbSet}}

\newtheorem{definition}{Definition}
\newcommand{\InInt}[1]{\mathsf{InInt(#1)}}

\algrenewcommand\algorithmicindent{.5em}%

\newcommand{\code}[1]{\textsf{\small #1}\xspace}
\newcommand{\spec}{\mathsf{spec}}

\newcommand{\liveness}{\mathsf{liveness}}
\input{abs}
\input{intro}
\input{prelim}
\input{problem}

\input{progression}

\input{smt}
\input{eval}

\section{Related Work}
\label{sec:related}

Centralized and decentralized online predicate detection in an asynchronous distributed system have been studies in~\cite{cgnm13,mg05}. Extensions to include temporal operators appear in~\cite{og07,mb15}. The line of work in~\cite{cgnm13,mg05,og07,mb15,svar04} considers a fully asynchronous system. A SMT-based predicate detection solution has been introduced in~\cite{vyktd17}. On the other hand, runtime monitoring for synchronous distributed system has been studied in~\cite{ds19,cf16,bf16}. This approach has shortcoming, the major one being the assumption of a common global clock shared among all the processes. Finally, fault-tolerant monitoring, where monitors can crash, has been investigated in~\cite{bfrrt16} for asynchronous and in~\cite{kb18} for synchronized distributed processes.

Runtime monitoring of time sensitive distributed system has been studied in~\cite{basin2015, 
basin2010, worrell2019, rosu2005}. With the onset of blockchains and the security vulnerability 
posed by smart contracts have been studied in~\cite{garcia2020, pace2021, pace2021a, rosu2018, 
rosu2018a}. The major area that these work lack is all of them consider the system to be 
synchronous with the presence of a global clock. However, smart contracts often include multiple 
blockchains and thus we consider a partially synchronous system where a synchronization algorithm 
limits the maximum clock skew among processes to a constant. An SMT-based solution was studied 
in~\cite{ganguly2020}, which we extend to include more expressive time bounded logic.

\section{Conclusion}
\label{sec:concl}

In this paper, we study distributed runtime verification. We propose a technique which takes an \MTL 
formula and a distributed computation as input. By assuming partial synchrony among all processes, 
first we chop the computation into several segments and then apply a progression-based formula 
rewriting monitoring algorithm implemented as a SMT decision problem in order to verify the 
correctness of the distributed system with respect to the formula. We conducted extensive synthetic 
experiments on trace generated by the \code{UPPAAL} tool and a set of blockchain smart contracts.

For future work, we plan to study the trade off among accuracy and scalability of our approach. 
Another important extension of our work is distributed runtime verification where the processes are 
dynamic, i.e., the process can crash and can also restore its state at any given time during 
execution. This will let us study a wide range of applications including airspace monitoring.

\bibliographystyle{IEEEtran}
\bibliography{bibliography}

\newpage
\input{appendix}

\end{document}

%% file: abs.tex
\begin{abstract}

Transactions involving multiple blockchains are implemented by {\em cross-chain} protocols.
These protocols are based on smart contracts,
programs that run on blockchains, executed by a network of computers.
Because smart contracts can automatically transfer ownership of cryptocurrencies,
electronic securities, and other valuable assets among untrusting parties,
verifying the runtime correctness of smart contracts is a problem of compelling practical interest.
Such verification is challenging since smart contract execution is time sensitive,
and the clocks on different blockchains may not be perfectly synchronized.
This paper describes a method for runtime monitoring of blockchain executions.
First, we propose a generalized runtime verification technique for verifying partially synchronous distributed computations for the metric temporal logic (\MTL) by exploiting bounded-skew clock synchronization.
Second, we introduce a progression-based formula rewriting scheme for monitoring \MTL specifications which employs SMT solving techniques and report experimental results.

\end{abstract}


%% file: intro.tex
\section{Introduction}
\label{sec:intro}

{\em Blockchain} technology~\cite{lu2019blockchain,nakamoto2008bitcoin} is a blockbuster in this era. 
It has drawn extensive attention from both industry and academia. With blockchain technology, people can trade in a peer-to-peer manner without mutually trusting each other, removing the necessity of a trusted centralized party.
The concept of decentralization appears extremely appealing, and the transparency, anonymity, and persistent storage provided by blockchain make it more attractive.
This revolutionary technology has triggered many applications in industry, ranging from cryptocurrency~\cite{herlihy2018atomic}, non-fungible tokens~\cite{herlihy2021cross}, internet of things\cite{christidis2016blockchains} to
health services \cite{xu2019healthchain}, etc.

Besides the huge success of cryptocurrencies known as blockchain 1.0, especially Bitcoin~\cite{nakamoto2008bitcoin}, blockchain 2.0, known as {\em smart contracts}~\cite{cong2019blockchain}, is also promising in many scenarios.
Smart contract is a program running on the blockchain. 
Its execution is triggered automatically and enforced by conditions preset in the code.
In this way, the transfer of assets can be automated by the rules in the smart contracts, and human intervention cannot stop it.
A typical smart contract implementation is provided by 
\code{Ethereum}~\cite{dannen2017introducing}, which uses {\em Solidity} 
\cite{dannen2017introducing}, which is a Turing-complete language.
However, automating the transactions by smart contracts also has its downsides.
If the smart contract has bugs and does not do what is expected, then lack of human intervention may lead to massive financial losses.
For example, as pointed out by~\cite{ellul2018runtime}, the Parity Multisig Wallet smart contract~\cite{parity} version 1.5 included a vulnerability which led to the loss of 30 million US dollars. %
Thus, developing effective techniques to verify the correctness of smart contracts is both urgent and important to protect against possible losses.
Furthermore, when a protocol is made up of multiple smart contracts across different blockchains, the correctness of protocols also need to be verified.

In this paper, we advocate for a {\em runtime verification} (RV) approach, to monitor the behavior of a system of blockchains with respect to a set of temporal logic formulas.
Applying RV to deal with multiple blockchains can be reduced to {\em distributed RV}, where a centralized or decentralize monitor observes the behavior of a distributed system in which processes do not share a global clock. 
Although RV deals with finite executions, the lack of a common global clock prohibits it from having a unique ordering of events in a distributed setting.
Put it another way, the monitor can only form a partial order of event which may result in different verification verdicts.
Enumerating all possible ordering of events at run time incurs in an exponential blow up, making the approach not scalable.
To add to this already complex task, most specification for verifying blockchain smart contracts, come with a time bound.
This means, not only the ordering of the events are at play when verifying, but also the actual physical time of occurrence of the event dictates the verification verdict.

\begin{figure}
    \centering
    \includegraphics[width=0.47\textwidth]{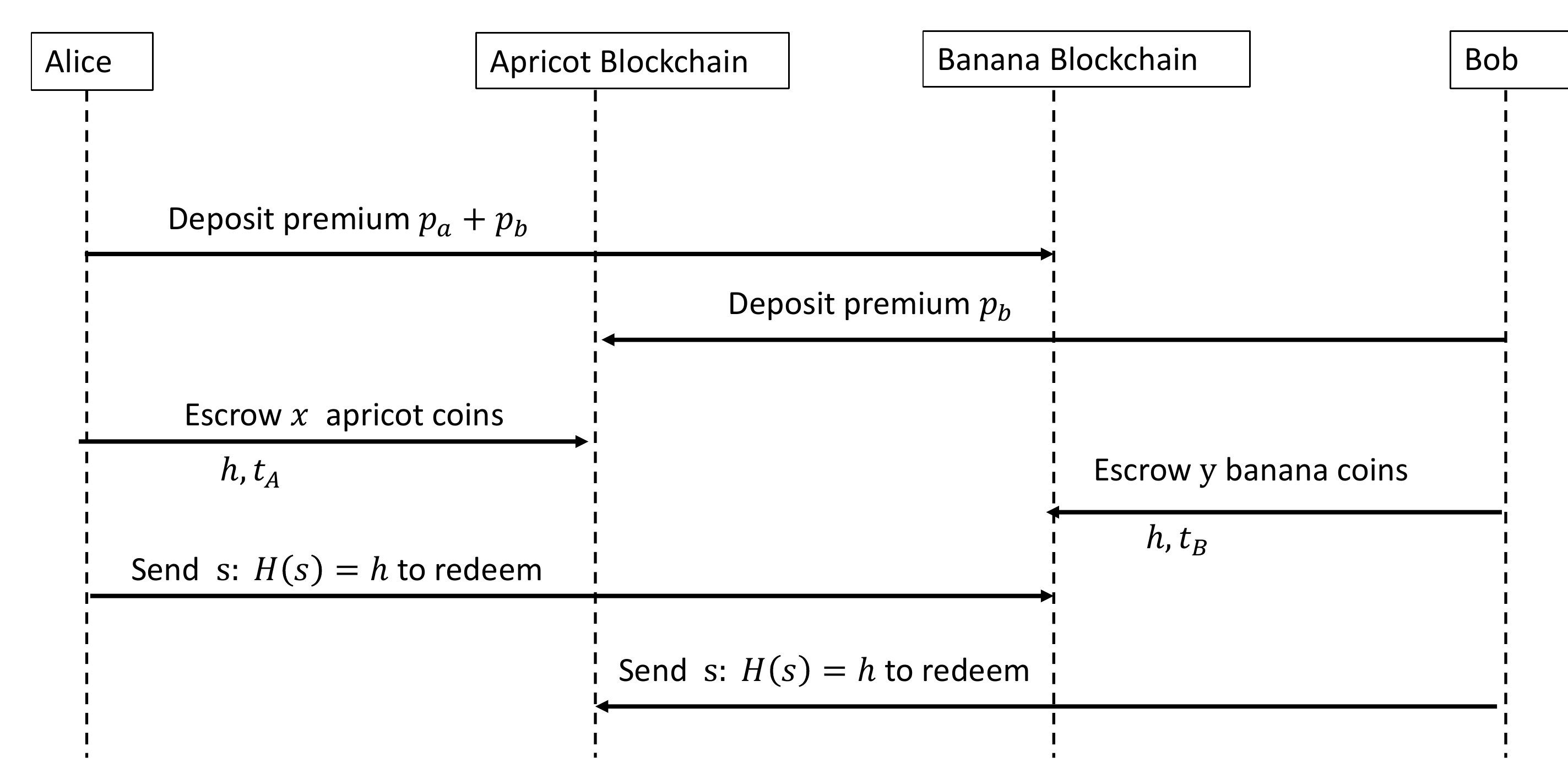}
    \caption{Hedged Two-party Swap}
    \label{fig:intro1}
    \vspace{-5mm}
\end{figure}

In this paper, we propose an effective, sound and complete solution to distributed RV for timed specifications expressed in the {\em metric temporal logic} (\MTL)~\cite{koy90}. 
To present a high-level view of \MTL, consider the {\em two-party swap} protocol~\cite{herlihy2021cross} shown in Fig~\ref{fig:intro1}.
Alice and Bob, each in possession of Apricot and Banana blockchain assets respectively, wants to swap their assets between each other without being a victim of sore-loser attack. 
There is a number of requirements that should be followed by conforming parties to discourage any attack on themselves.
We use the {\em metric temporal logic} (\MTL)~\cite{koy90} to express such requirements. One such requirement, where Alice should not redeem her asset before Bob within eight time units can be represented by the \MTL formula:
$$
\varphi_{\spec} = \neg \texttt{Apr.Redeem}(\mathit{bob}) \, \U_{[0, 8)} \texttt{Ban.Redeem}(\mathit{alice}).
$$

We consider a fault proof central monitor which has the complete view of the system but has no access to a global clock.
In order to limit the blow-up of states posed by the absence of a global clock, we make a practical assumption about the presence of a {\em bounded clock skew} $\epsilon$ between the local clocks of every pair of processes.
This is guaranteed by a synchronization algorithm (e.g. NTP~\cite{ntp}).
This setting is known to be partially synchronous when we do not assume any presence of a global clock and limit the impact of asynchrony within clock drifts. 
Such an assumption limits the window of partial orders of events only within $\epsilon$ time units and significantly reduces the combinatorial blow-up caused by nondeterminism due to concurrent. 
Existing distributed RV techniques either assume a global clock when working with time sensitive specifications~\cite{basin2015, worrell2019} or use untimed specifications when assuming partial synchrony~\cite{ganguly2020,mbab21}.

\input{fig_intro2}

We introduce an SMT\footnote{{\em Satisfiability modulo theories} (SMT) is the problem of determining whether a formula involving Boolean expressions comprising of more complex formulas involving real numbers, integers, and/or various data structures is satisfiable.}-based {\em progression-based} formula rewriting technique over distributed computations which takes into consideration the events observed thus far to rewrite the specifications for future extensions.
Our monitoring algorithm accounts for all possible orderings of events without explicitly generating them when evaluating \MTL formulas.
For example, in Fig.~\ref{fig:intro2}, we see the events and the time of occurrence in the 
two blockchains, Apricot($Apr$) and Banana($Ban$) divided into two segments, $seg_1$ and 
$seg_2$ for computational purposes. Considering maximum clock skew $\epsilon = 2$ and the 
specification $\varphi_{\spec}$, at the end of the first segment, we have two possible rewritten 
formulas for the next segment:
\begin{align*}
\varphi_{\spec_1} & = \neg \texttt{Apr.Redeem}(bob) \, \U_{[0, 4)} \texttt{Ban.Redeem}(alice)\\
%
\varphi_{\spec_2} & = \neg \texttt{Apr.Redeem}(bob) \,  \U_{[0, 3)} \texttt{Ban.Redeem}(alice)
\end{align*}
This is possible due to the different ordering and different time of occurrence of the events $\texttt{Deposit}(p_b)$ and $\texttt{Deposit}(p_a + p_b)$.
In other words, the possible time of occurrence of the event $\texttt{Deposit}(p_b)$ (resp. $\texttt{Deposit}(p_a + p_b)$) is either 2, 3 or 4 (resp. 3, 4, or 5) due to the maximum clock skew of 2.
Likewise, at the end of $seg_2$, we have $\varphi_{\spec_1}$ evaluate to $\tru$ where as 
$\varphi_{\spec_2}$ evaluate to $\fals$.
This is because, even if we consider the scenario when $\texttt{Ban.Redeem(alice)}$ occurs before 
$\texttt{Apr.Redeem(bob)}$, a possible time of occurrence of $\texttt{Ban.Redeem(alice)}$ is $8$ 
(resp. $6$) which makes $\varphi_{\spec_2}$ 
(resp. $\varphi_{\spec_1}$) evaluate to $\fals$  (resp. $\tru$).

We have fully implemented our technique\footnote{\url{https://github.com/ritam9495/rv-mtl-blockc}} and report the results of rigorous experiments on monitoring synthetic data, using benchmarks in the tool \code{UPPAAL}~\cite{lpy97}, as well as monitoring correctness, liveness and conformance conditions for smart contracts on blockchains. We put our monitoring algorithm to test studying the effect of different parameters on the runtime and report on each of them. Using our technique we learn not to use a value of $\Delta$ (transaction deadline) that is comparable to the value of clock skew $\epsilon$ when designing the smart contract.

\paragraph*{Organization} Section~\ref{sec:prelim} presents the background concepts. Formal statement of our RV problem  is discussed in Section~\ref{sec:sol}. The formula progression rules and the SMT-based solution are described in Sections~\ref{sec:progress} and~\ref{sec:smt}, respectively, while experimental results are analyzed in Section~\ref{sec:eval}. Related work is discussed in Section~\ref{sec:related} before we make concluding remarks in Section~\ref{sec:concl}. The appendix includes more details about our case studies.

%% file: fig_intro2.tex
\begin{figure}
    \centering 
    \scalebox{0.65}{
        \begin{tikzpicture}
        
            \draw (-0.5, 2) node[] {$Apr$};
            \draw (-0.5, 0) node[] {$Ban$};
            
            \draw [->] (0,2) -- (4,2);
            \draw [->] (0,0) -- (4,0);
            
            \draw [fill = white] (1.1,2) circle (0.1) node[above, yshift=0.1cm]{$\texttt{SetUp}$};
            \draw (1.1,2) node[below, yshift=-0.1cm]{$1$};
            \draw [fill = white] (3,2) circle (0.1) node[above, yshift=0.1cm]{$\texttt{Deposit}(p_b)$};
            \draw (3,2) node[below, yshift=-0.1cm]{$3$};
            
            \draw [fill = white] (1,0) circle (0.1) node[above, yshift=0.1cm]{$\texttt{SetUp}$};
            \draw (1,0) node[below, yshift=-0.1cm]{$1$};
            \draw [fill = white] (3.5,0) circle (0.1) node[above, yshift=0.1cm]{$\texttt{Deposit}(p_a + p_b)$};
            \draw (3.5,0) node[below, yshift=-0.1cm]{$4$};
            
            \draw [blue, dashed] plot [smooth, tension=0.5] coordinates {(2.8,2.25) (2.7,1.75) (3.8,0.25) (3.7,-0.25)};
            \draw [red, dashed] plot [smooth, tension=0.5] coordinates {(3.2,2.25) (3.3,1.75) (3.1,0.25) (3.2,-0.5)};
            
            \draw (5.5,2) node[] {$Apr$};
            \draw (5.5,0) node[] {$Ban$};
            
            \draw [->] (6,2) -- (10,2);
            \draw [->] (6,0) -- (10,0);
            
            \draw [fill = white] (6.5,2) circle (0.1) node[above, yshift=0.1cm]{$\texttt{Escrow}(h, t_A)$};
            \draw (6.5,2) node[below, yshift=-0.1cm]{$5$};
			\draw [fill = white] (9.4,2) circle (0.1) node[above, yshift=0.1cm]{$\texttt{Redeem}(bob)$};
			\draw (9.4,2) node[below, yshift=-0.1cm]{$7$};
            
            \draw [fill = white] (7,0) circle (0.1) node[above, yshift=0.1cm]{$\texttt{Escrow}(h, t_B)$};
            \draw (7,0) node[below, yshift=-0.1cm]{$6$};
            \draw [fill = white] (9.5,0) circle (0.1) node[above, yshift=0.1cm]{$\texttt{Redeem}(alice)$};
            \draw (9.5,0) node[below, yshift=-0.1cm]{$7$};
            
            \draw [blue, dashed] plot [smooth, tension=0.5] coordinates {(9.2,2.25) (9,1.75) (9.9,0.25) (9.7,-0.25)};
            \draw [red, dashed] plot [smooth, tension=0.5] coordinates {(9.6,2.25) (9.8,1.75) (9.1,0.25) (9.3,-0.5)};
            
            \draw (2,-0.5) node[] {${seg}_1$};
            \draw (7.5,-0.5) node[] {${seg}_2$};
            
        \end{tikzpicture}
    }
    \caption{Progression Example}
    \label{fig:intro2}
    \vspace{-5mm}
\end{figure}
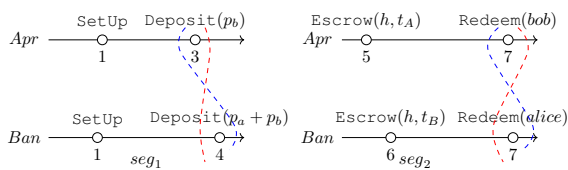

%% file: prelim.tex
\section{Preliminaries}
\label{sec:prelim}

In this section, we  present an overview of the distributed computations and the metric temporal logic (\MTL).

\subsection{Distributed Computation}

We consider a loosely coupled asynchronous message passing system, consisting of $n$ reliable 
processes (that do not fail), denoted by $\Proc = \{ P_1, P_2, \cdots, P_n \}$. As a system, the 
processes do not share any memory or have a common global clock. Channels are assumed to be 
FIFO and lossless. In our model, we represent each local state change by an event and a message 
activity (send or receive) is represented by an event as well. Message passing does not change the 
state of the process and we disregard the content of the message as it is of no use for our 
monitoring technique. Here, we refer to a global clock which will act as the ``real" timekeeper. It is to 
be noted that the presence of this global clock is just for theoretical reasons and it is not available to 
any of the individual processes.

We make an assumption about a partially synchronous system. For each process $P_i$, where $i \in [1, n]$, the local clock can be represented as a monotonically increasing function $c_i: \wholePlusSet \rightarrow \wholePlusSet$, where $c_i(\globalC)$ is the value of the local clock at global time $\globalC$. Since we are dealing with discrete-time systems, for simplicity and without loss of generality, we represent time with non-negative integers $\wholePlusSet$. For any two processes $P_i$ and $P_j$, where $i \neq j$, we assume:
$$
\forall \globalC \in \wholePlusSet. \mid c_i(\globalC) - c_j(\globalC) \mid < \epsilon,
$$
where $\epsilon > 0$ is the maximum clock skew. The value of $\epsilon$ is constant and is known to the monitor. This assumption is met by the presence of a clock synchronization algorithm, like NTP~\cite{ntp}, to ensure bounded clock skew among all processes.

We denote an {\em event} on process $P_i$ by $e^i_{\RTime}$, where $\RTime = c_i(\globalC)$, that is the local time of occurrence of the event at some global time $\globalC$.
%

\begin{definition}
A {\em distributed computation} consisting of $n$ processes is represented by the pair $(\Events, 
 \hb)$, where
$\Events$ is a set of events partially ordered by Lamport's 
happened-before ($\hb$) relation~\cite{hb1978}, subject to the partial synchrony assumption:

\begin{itemize}

    \item For every process $P_i$, $1 \leq i \leq n$, all the events happening on it are totally ordered, 
    that is, 
    $$\forall \RTime, \RTime' \in \wholePlusSet: (\RTime < \RTime') \rightarrow (e^i_{\RTime} \hb e^i_{\RTime'});
    $$

    \item If $e$ is a message sending event in a process and $f$ is the corresponding message 
    receiving event in another process, then we have $e \hb f$;

    \item For any two processes $P_i$ and $P_j$ and two corresponding events $e^i_{\RTime}, e^j_{\RTime'} \in 
    \Events$, if $\RTime + \epsilon < \RTime'$ then, $e^i_{\RTime} \hb e^j_{\RTime'}$, where $\epsilon$ is 
    the maximum clock skew, and

    \item If $e \hb f$ and $f \hb g$, then $e \hb g$.\qed
\end{itemize}
\end{definition}

\begin{definition}
Given a distributed computation $(\Events, \hb)$, a subset of events $\cc \subseteq 
\Events$ is said to form a consistent cut if and only if when $\cc$ contains an event $e$, then it should 
also contain all such events that happened before $e$. Formally, 
$$
\forall e \in \Events. (e \in \cc) \land (f \hb e) \rightarrow f \in \cc.~\blacksquare
$$
\end{definition}

The frontier of a consistent cut $\cc$, denoted by $\front(\cc)$ is the set of all events that happened last in each process in the cut. That is, $\front(\cc)$ is a set of $e^i_{last}$ for each $i \in [1, |\Proc|]$ and $e^i_{last} \in \cc$. We denote $e^i_{last}$ as the last event in $P_i$ such that $\forall e^i_{\RTime} \in \cc. (e^i_{\RTime} \neq e^i_{last}) \rightarrow (e^i_{\RTime} \hb e^i_{last})$.

\subsection{Metric Temporal Logic (MTL)~\cite{mtl1992,mtl1994}}
\label{subsec:mtl}

Let $\intervalSet$ be a set of nonempty intervals over $\wholePlusSet$. We define an interval, $\interval$, to be 
$$
[\Istart, \Iend) \triangleq \{ a \in \wholePlusSet \mid \Istart \leq a < \Iend \}
$$
where $\Istart \in \wholePlusSet$, $\Iend \in \wholePlusSet \cup 
\{ \infty \}$ and $\Istart < \Iend$. We define $\Pred$ as the set of all {\em atomic propositions} and $\Sigma = 2^{\Pred}$ 
as the set of all possible {\em states}.
A {\em trace} is represented by a pair which consists of a sequence of states, denoted by $\trace = 
s_0s_1 \cdots$, where $s_i \in \Sigma$ for every $i > 0$ and a sequence of non-negative numbers, 
denoted by $\TimeV = \Time_0\Time_1 \cdots$, where 
$\Time_i \in \wholePlusSet$ for all $i > 0$. We represent the set of all infinite traces by a pair of 
infinite sets, $(\Sigma^\omega, \wholePlusSet^\omega)$. The trace $s_ks_{k+1}\cdots$ (resp. 
$\Time_k\Time_{k+1}$) is represented by $\trace^k$ (resp. $\Time^k$). For an infinite trace $\trace = 
s_0s_1 \cdots$ and $\TimeV = \Time_0\Time_1 \cdots$, $\TimeV$ is a increasing
sequence, meaning $\Time_{i+1} \geq \Time_{i}$, for all $i \geq 0$.

\paragraph*{Syntax.} The syntax of metric temporal logic (\MTL) for infinite traces are defined by the following grammar:
$$
\varphi ::= p \mid \neg \varphi \mid \varphi_1 \lor\varphi_2 \mid \varphi_1 \U_\interval \varphi_2
$$
where $p \in \Pred$ and $\U_\interval$ is the `until' temporal operator with time interval $\interval$.
Note that other propositional and temporal operators can be represented using the ones mentioned above. For example, $\tru = p \lor \neg p$, $\fals = \neg \tru$, $\varphi_1 \rightarrow \varphi_2 = \neg \varphi_1 \lor \varphi_2$, $\varphi_1 \land \varphi_2 = \neg (\neg \varphi_1 \lor \neg \varphi_2)$, $\F_\interval \varphi = \tru \U_\interval \varphi$ (``eventually") and $\G_\interval \varphi = \neg (\F_\interval \neg \varphi)$ (``always"). We denote the set of all \MTL formulas by $\Phi_{\MTL}$.

\paragraph*{Semantics}The semantics of metric temporal logic (\MTL) is 
defined over the trace, $\trace = s_0s_1 \cdots$ and $\TimeV = \Time_0\Time_1 \cdots$ as follows:
\[
\begin{array}{l l l}
	(\trace, \TimeV, i) \models p~ & \text{iff} & p \in s_i \\
	(\trace, \TimeV, i) \models \neg \varphi & \text{iff} & (\trace, \TimeV, i) \not\models \varphi \\
	(\trace, \TimeV, i) \models \varphi_1 \lor \varphi_2 & \text{iff} & (\trace, \TimeV, i) \models \varphi_1 
	\lor (\trace, \TimeV, i) \models \varphi_2 \\
	(\trace, \TimeV, i) \models \varphi_1 \U_\interval \varphi_2 & \text{iff} & \exists j \geq i.
	\Time_j - \Time_i \in \interval \land (\trace, \TimeV, j) \models \\
	& & \varphi_2 \land \forall k \in [i, j), (\trace, \TimeV, k) \models \varphi_1
\end{array}
\]
It is to be noted that $(\trace, \TimeV) \models \varphi$ holds if and only if $(\trace, \TimeV, 0) \models \varphi$.

In the context of runtime verification, we introduce the notion of finite \MTL. The truth values
are represented by the set $\mathsf{B}_2 = \{ \top, \bot \}$, where $\top$ (resp. $\bot$) 
represents a formula that is satisfied (resp. violated) given a finite trace. 
We represent the set of all finite traces by a pair of finite sets, $(\Sigma^{*}, 
\wholePlusSet^{*})$.
For a finite trace, $\trace = s_0s_1\cdots s_n$ and 
$\TimeV = \Time_0\Time_1\cdots \Time_n$ the only semantic that needs to be redefined is that of
$\U$ (`until') and is as follows:
\begin{equation}
	\nonumber
	[(\trace, \TimeV, i) \models_F \varphi_1 \U_\interval \varphi_2] = \begin{cases}
		\top & \text{if }\exists j \geq i. \Time_j - \Time_i \in \interval \\
		& ~([\trace^j \models_F \varphi_2] = \top) \land \forall k \in \\
		& ~[i, j) : ([\trace^k \models_F \varphi_1] = \top)\\
		\bot & \text{otherwise}.
	\end{cases}
\end{equation}

In order to further illustrate the difference between \MTL and finite \MTL, we consider the formula $\varphi = \F_\interval p$ and a trace $\trace = s_0s_1\cdots s_n$ and $\TimeV = \Time_0\Time_1\cdots\Time_n$. We have $[(\trace, \TimeV) \models_F \varphi] = \top$ if for some $j \in [0, n]$ we have $\Time_j - \Time_0 \in \interval$ and $p \in s_i$, otherwise $\bot$. Now, consider the formula $\varphi = \G_\interval p$ we have $[(\trace, \TimeV) \models_F \varphi] = \bot$ if for some $j \in [0, n]$ we have $\Time_j - \Time_0 \in \interval$ and $p \not\in s_i$, otherwise $\top$.

%% file: problem.tex
\section{Formal Problem Statement}
\label{sec:sol}

In a partially synchronous system, there are different ordering of events that is possible and each 
unique {\em ordering} of events~\cite{ylies2012} might evaluate to different verdicts. In other words, 
a partially synchronous distributed computation $(\Events, \hb)$ may have different ordering of 
events primarily due to the different interleavings of events that is possible. Thus, it is possible to 
have different verdicts for the same distributed computation for the different ordering of events. 

Let $(\Events, \hb)$ be a distributed computation. A sequence of consistent cuts is of the form 
$\cc_0\cc_1\cc_2 \cdots$, where for all $i \geq 0$, we have (1) $\cc_i \subset \cc_{i+1}$ and (2) 
$|\cc_i| + 1 = |\cc_{i+1}|$, and (3) $\cc_0 = \emptyset$. The set of all sequences of consistent cuts 
be denoted by $\ccAll$.
We note that we view time interval $\interval$ in the syntax of \MTL is in terms of the physical 
(global) time $\globalC$. Thus, when deriving all the possible traces given the distributed 
computation \linebreak $(\Events, \hb)$, we have to account for all different orders in which the 
events could possibly had occur with respect to $\globalC$. This involves replacing the local time of 
occurrence of an event, $e^i_\RTime$ with the set of event $\{e^i_{\RTime'} \mid \RTime' \in 
[\max\{0, \RTime - \epsilon + 1\}, \RTime + \epsilon)\}$. This is to account for the maximum clock drift 
that is possible on the local clock of a process when compared to the global clock.
\input{fig_prelim_ps}
For example, given the computation in Figure~\ref{fig:prelim}, a maximum clock skew $\epsilon = 2$ 
and a \MTL formula, $\varphi = a \U_{[0, 6)} b$, one has to consider all possible traces including 
$(a,1)(a,2)(b,4)(\neg a,5) \models \varphi$ and $(a,1)(a,2)(\neg a,4)(b,5) \not\models \varphi$. The 
contradictory result is due to the different time of occurrence of event that needs to be considered.

Given a sequence of consistent cuts, it is evident that for all $j > 0$, $|\cc_j - \cc_{j-1}| = 1$ and 
event $\cc_j - \cc_{j-1}$ is the last event that was added onto the cut  $\cc_j$. To translate 
monitoring of a distributed system into monitoring a trace, We define a sequence of natural numbers 
as $\LTimeV = \LTime_0\LTime_1\cdots$, where $\LTime_0 = 0$ and for each $j \geq 1$, we have 
$\LTime_{j} = \RTime$, such that $\front(\cc_{j}) - \front(\cc_{j-1}) = \{e^i_{\RTime}\}$. To maintain 
time monotonicity, we only consider sequences where for all $i \geq 0$, $\LTime_{i+1} \geq 
\LTime_i$.

The set of all traces that can be formed from $(\Events, \hb)$ is defined as:
$$
\Tr(\Events, \hb) = \Big\{ \front(\cc_0)\front(\cc_1)\cdots \mid \cc_0\cc_1\cdots \in \ccAll \Big\}.
$$
In the sequel, we assume that every sequence $\alpha$ of frontiers in $\Tr(\Events, \hb)$ is 
associated with a sequence $\LTimeV$. Thus, to comply with the semantics of \MTL, we refer to the 
elements of $\Tr(\Events, \hb)$ by pairs of the form $(\trace, \LTimeV)$.
Now that we have a set of all possible traces, we evaluate an \MTL formula $\varphi$ with respect to 
the computation $(\Events, \hb)$ as follows:
$$
[(\Events, \hb) \models_F \varphi] = \Big\{ (\trace, \LTimeV, 0) \models_F \varphi \mid (\trace, 
\LTimeV) \in \Tr(\Events, \hb) \Big\}.
$$

This boils down to having a set of verdicts, since a distributed computation may involve several 
traces and each trace might evaluate to a different verdict.

\paragraph*{Overall Idea of our solution.} To solve the above problem (evaluating all possible 
verdicts), we propose a monitoring approach based on formula-rewriting
(Section~\ref{sec:progress}) and SMT solving (Section~\ref{sec:smt}).
Our approach involves iteratively(1) chopping a distributed computation into a sequence of smaller 
segments to reduce the problem size and (2) progress the \MTL formula for each segment for the 
next segment, which results in a new \MTL formula by invoking an SMT solver.
Since each computation/segment corresponds to a set of possible traces due to partial synchrony, 
each invocation of the SMT solver may result in a different verdict.

%% file: fig_prelim_ps.tex
\begin{figure}
    \centering 
    \scalebox{1}{
        \begin{tikzpicture}
        
            \draw (-0.5, 1) node[] {$P_1$};
            \draw (-0.5, 0) node[] {$P_2$};
            
            \draw [->] (0,1) -- (5,1);
            \draw [->] (0,0) -- (5,0);
            
            \draw [fill = white] (1.5,1) circle (0.2) node{$a$};
            \draw (1.5,1) node[below, yshift=-0.15cm]{$1$};
            \draw [fill = white] (4.25,1) circle (0.2) node{$\neg a$};
            \draw (4.25,1) node[below, yshift=-0.15cm]{$4$};
            
            \draw [fill = white] (1,0) circle (0.2) node{$a$};
            \draw (1,0) node[below, yshift=-0.15cm]{$2$};
            \draw [fill = white] (3.75,0) circle (0.2) node{$b$};
            \draw (3.75,0) node[below, yshift=-0.15cm]{$5$};
            
            \draw [blue, dashed] plot [smooth, tension=0.5] coordinates {(4.1,1.25) (3.9,0.75) (4.05,0.25) (3.95,-0.25)};
            \draw [red, dashed] plot [smooth, tension=0.5] coordinates {(4.4,1.25) (4.6,0.75) (3.45,0.25) (3.55,-0.25)};
            
        \end{tikzpicture}
    }
    \caption{Trace Example}
    \label{fig:prelim}
\end{figure}
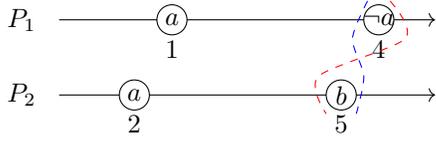

%% file: progression.tex
\section{Formula Progression for MTL}
\label{sec:progress}

We start describing our solution by explaining the formula progression technique.

\begin{definition}
	A {\em progression function} is of the form $\Pro : \Sigma^{*} \times \wholePlusSet^{*} \times \Phi_{\MTL} \rightarrow \Phi_{\MTL}$ and is defined for all finite traces $(\trace, \TimeV) \in 
	(\Sigma^{*}, \wholePlusSet^{*})$, infinite traces $(\trace', \TimeV') \in (\Sigma^{\omega},
	\wholePlusSet^{\omega})$ and \MTL formulas $\varphi \in \Phi_{\MTL}$, such that $(\trace.\trace', \TimeV.\TimeV') \models \varphi$ if and only if 
	$(\trace', \TimeV') \models \Pro(\trace, \TimeV, \varphi)$. \qed
\end{definition}
\vspace{2mm}

It is to be noted that compared to the classic formula regression technique in~\cite{Klaus2001}, here 
the function $\Pro$ takes a finite trace as input, while the algorithm in~\cite{Klaus2001} rewrite the 
formula after every observed state. When monitoring a partially synchronous distributed system, 
where multiple verdicts are possible and no unique ordering of events are possible, the classical 
state-by-state formula rewriting technique is of little use. The motivation of our approach comes 
from the fact that for computation reasons, we chop the computation into smaller segments 
and the verification of each segment is done through an SMT query. A state-by-state approach would 
incur in a huge number of SMT queries being generated.


Let $\interval = [\Istart, \Iend)$ denote an interval. By $\interval - \Time$, we mean the interval $\interval' = [\Istart', \Iend')$, where $\Istart' = \max\{0, \Istart - \Time \}$ and $\Iend' = \max\{0, \Iend - \Time \}$.
Also, for two time instances, $\tau_i$ and $\tau_0$, we let $\InInt{i}$ return $\tru$ or $\fals$ depending upon the whether $\tau_i - \tau_0 \in \interval$.

\begin{figure*}
\begin{minipage}{0.48\textwidth}
    \input{alg_globally}
\end{minipage}
\hfill
\begin{minipage}{0.48\textwidth}
    \input{alg_eventually}
\end{minipage}
\end{figure*}

\input{alg_until}

\vspace{2mm}
\noindent \textbf{Progressing atomic propositions.} For an \MTL formula of the form $\varphi = p$, where $p \in \Pred$, the result depends on whether or not $p \in \trace(0)$. This marks as our base case for the other temporal and logical operators:
\begin{equation}
	\nonumber
	\Pro(\trace, \TimeV, \varphi) = 
	\begin{cases} 
		\tru & \text{if }~p \in \trace(0) \\
		\fals & \text{if }~p \not\in \trace(0)
	\end{cases}
\end{equation}

\noindent \textbf{Progressing negation.} For an \MTL formula of the form $\varphi = \neg \phi$, we have: 
$$
\Pro(\trace, \TimeV, \varphi) = \neg  \Pro(\trace, \TimeV, \phi).
$$

\noindent \textbf{Progressing disjunction.} Let $\varphi = \varphi_1 \lor \varphi_2$. Apart from the trivial cases, the result of progression of $ \varphi_1 \lor \varphi_2$ is based on progression of $\varphi_1$ and/or progression of $\varphi_2$: 
\begin{equation}
	\nonumber
	\Pro(\trace, \TimeV, \varphi) = 
	\begin{cases} 
		\tru & \text{if }~\Pro(\trace, \TimeV, \varphi_1) = \tru \: \lor \\
		& \Pro(\trace, \TimeV, \varphi_2) = \tru \\
		\fals & \text{if }~\Pro(\trace, \TimeV, \varphi_1) = \fals \: \land \\
		& \Pro(\trace, \TimeV, \varphi_2) = \fals \\
		\varphi_2' & \text{if }~\Pro(\trace, \TimeV, \varphi_1) = \fals \: \land \\
		& \Pro(\trace, \TimeV, \varphi_2) = \varphi_2' \\
		\varphi_1' & \text{if }~\Pro(\trace, \TimeV, \varphi_2) = \fals \: \land \\
		& \Pro(\trace, \TimeV, \varphi_1) = \varphi_1' \\
		\varphi_1' \lor \varphi_2' & \text{if }~\Pro(\trace, \TimeV, \varphi_1) = \varphi_1' \: \land \\
		& \Pro(\trace, \TimeV, \varphi_2) = \varphi_2'
	\end{cases}
\end{equation}


\noindent \textbf{Always and eventually operators.} As shown in Algorithms~\ref{alg:globally} and 
~\ref{alg:eventually}, the progression for `always', $(\G_\interval \varphi)$ and `eventually', $(\F_\interval \varphi)$ depends on the value of $\InInt{i}$ and the progression of the inner formula $\varphi$. In Algorithm~\ref{alg:globally} and \ref{alg:eventually}, we divide the algorithm into three cases: (1) line 4, corresponds to if the $\interval$ is within the sequence $\TimeV$; (2) line 6, corresponds to where $\interval$ starts in the current trace but its end is beyond the boundary of the sequence $\TimeV$, and (3) line 9, corresponds to if the entire interval $\interval$ is beyond the boundary of sequence $\TimeV$. 
In Algorithm~\ref{alg:globally}, we are only concerned about the progression of $\varphi$ on the suffix  $(\trace^i, \TimeV^i)$ if $\InInt{i} = \tru$. 
In case, $\InInt{i} = \fals$ the consequent drops and the entire condition equates to $\tru$. In other words, equating over all $i \in [0, |\trace|]$, we are only left with conjunction of $\Pro(\trace^i, \TimeV^i, \varphi)$ where $\InInt{i} = \tru$. In addition to this, we add the initial formula with updated interval for the next trace. Similarly, in Algorithm~\ref{alg:eventually}, equating over all $i \in [0, |\trace|]$, if $\InInt{i} = \fals$ the corresponding $\Pro(\trace^i, \TimeV^i, \varphi)$ is disregarded and the final formula is a disjunction of $\Pro(\trace^i, \TimeV^i, \varphi)$ with $\InInt{i} = \tru$.

\vspace{2mm}
\noindent \textbf{Progressing the until operator.} Let the formula be of the form $\varphi_1 \U_\interval \varphi_2$.
According to the semantics of until $\varphi_1$ should be evaluated to true in all states leading up to some $i \in \interval$, where $\varphi_2$ evaluates to true. We start by progressing $\varphi_1$ (resp. $\varphi_2$) as $\G_{[0, \Time_i - \Time_0)} \varphi_1$ (resp. $\F_{[\Time_i, \Time_i + 1)} \varphi_2$) for some $i \in \interval$.
Since, we are only verifying the sub-formula, $\F_{[\Time_i, \Time_i + 1)} \varphi_2$, on the trace sequence $(\trace, \TimeV)$, it is equivalent to verifying the sub-formula $\F_{[0, 1)} \varphi_2 \equiv \varphi_2$ over the trace sequence $(\trace^i, \TimeV^i)$.
Similar to Algorithms~\ref{alg:globally} and~\ref{alg:eventually}, in Algorithm~\ref{alg:until} we need to consider three cases. 
In lines 4, 6 and 9, following the semantics of until operator, we make sure for all $i \in [0, |\trace|]$, if $\Time_i < \interval_{start} + \Time_0$, $\varphi_1$ is satisfied in the suffix $(\trace^i, \TimeV^i)$. In addition to this there should be some $j \in [0, |\trace|]$ for which if $\InInt{j} = \tru$, then the trace satisfies the sub-formula $\G_{[0, \Time_j - \Time_0)} \varphi_1$ and $\F_{[\Time_j, \Time_j + 1)} \varphi_2$).
In lines 6 and 9, we also accommodate for future traces satisfying the formula $\varphi_1 \U_\interval \varphi_2$ with updated intervals.

\input{smt_prog_example}

\paragraph*{Example} In Fig.~\ref{fig:progression}, the time line shows propositions and their time of
occurrence, for formula $\F_{[0,6)} r \rightarrow ( \neg p \U_{[2, 9)} q)$. The entire computation 
has been divided into 3 segments, $(\trace, \TimeV)$, $(\trace', \TimeV')$, and $(\trace'', \TimeV'')$ 
and each state has been represented by $(s, \Time)$:

\begin{itemize}
	\item We start with segment $(\trace, \TimeV)$.
	First we evaluate $\F_{[0,6)} r$, which requires evaluating
	$\Pro(\trace^i, \TimeV^i, r)$ for $i \in \{0, 1, 2\}$, all of which returns the verdict $\fals$ and there 
	by rewriting the
	sub-formula as $\F_{[0, 4)} r$. Next, to evaluate the sub-formula $\neg p \U_{[2, 9)} q$, we need 
	to evaluate 
	(1) $\Pro(\trace^i, \TimeV^i, \neg p)$ for $i \in \{0, 1\}$ since $\Time_i - \Time_0 < 2$ and both 
	evaluates to $\tru$, 
	(2) $\Pro(\trace, \TimeV, \G_{[0, 2)} \neg p)$ which also evaluates to $\tru$ and 
	(3) $\Pro(\trace^2, \TimeV^2, q)$ which evaluates as $\fals$. Thereby, the rewritten formula
	after observing $(\trace, \TimeV)$ is $\F_{[0, 3)} r \rightarrow (\neg p \U_{[0, 6)} q)$.
	
	\item Similarly, we evaluate the formula now with respect to $(\trace', \TimeV')$, which makes the 
	sub-formula 
	$\F_{[0, 3)} r$ evaluate to $\tru$ at $\Time = 3$ and the sub-formula $\neg p \U_{[0, 6)} q$ 
	(there is no such $i \in \{0, 1, 2\}$ where $\Time_i - \Time_0 < 0$ and for all $j \in \{0, 1, 2\}$, 
	$\Pro(\trace'^j, \TimeV'^j, q) = \fals$) is rewritten as $\neg p \U_{[0, 4)} q$. 
	
	\item In $(\trace'', \TimeV'')$, 
	for $j = 1$, $\Pro(\trace'', \TimeV'', \G_{[0, 2)} \neg p) = \tru$ and $\Pro(\trace''^j, \TimeV''^j, q) = 
	\tru$, and thereby rewriting the entire formula as $\tru$.
\end{itemize}

%% file: alg_globally.tex
\begin{algorithm}[H]
	\caption{Always}\label{alg:globally}
	\begin{algorithmic}[1]
	\footnotesize
		\Function{$\Pro$}{$\trace, \TimeV, \G_\interval \varphi$}
		\If{$\interval_{start} \leq \Time_{|\trace|} - \Time_0$}
		    \If{$\interval_{end} \leq \Time_{|\trace|} - \Time_0$}
    		    \State \textbf{return} $\bigwedge _{i \in [0, |\trace|]} \big(\InInt{i} 
    		    \rightarrow \Pro(\trace^i, \TimeV^i, \varphi) \big)$
    		\Else
    		    \State \textbf{return} $\bigwedge_{i \in [0, |\trace|]} \big(\InInt{i} \rightarrow 
    		    \Pro(\trace^i, \TimeV^i, \varphi) \big) \land \G_{[\interval - (\Time_{|\trace|} - 
    		    \Time_0 ))} \varphi$
		    \EndIf
		\Else
		    \State \textbf{return} $\G_{[\interval -(\Time_{|\trace|} - \Time_0))} \varphi$
		\EndIf
		\EndFunction
	\end{algorithmic}
\end{algorithm}

%% file: alg_eventually.tex
\begin{algorithm}[H]
	\caption{Eventually}\label{alg:eventually}
	\begin{algorithmic}[1]
		\footnotesize
		\Function{$\Pro$}{$\trace, \TimeV, \F_\interval \varphi$}
		\If{$\interval_{start} \leq \Time_{|\trace|} - \Time_0$}
		    \If{$\interval_{end} \leq \Time_{|\alpha|} - \Time_0$}
		        \State \textbf{return} $\bigvee_{i \in [0, |\trace|]} \big(\InInt{i} \land 
		        \Pro(\trace^i, \TimeV^i, \varphi)\big)$
		    \Else
		        \State \textbf{return} $\bigvee_{i \in [0, |\trace|]} \big(\InInt{i} \land
		        \Pro(\trace^i, \TimeV^i, \phi) \big) \lor \F_{[\interval -(\Time_{|\trace|} - \Time_0))} \varphi$
		    \EndIf
		\Else
		    \State \textbf{return} $\F_{[\interval -(\Time_{|\trace|} - \Time_0))} \varphi$
		\EndIf
	    \EndFunction
	\end{algorithmic}
\end{algorithm}

%% file: alg_until.tex
\begin{algorithm*}
	\caption{Until}\label{alg:until}
	\begin{algorithmic}[1]
		\footnotesize
		\Function{$\Pro$}{$\trace, \TimeV, \varphi_1 \U_\interval \varphi_2$}
		\If{$\interval_{start} \leq \Time_{|\trace|}  - \Time_0$}
		    \If{$\interval_{end} \leq \Time_{|\trace|} - \Time_0$}
		        \State \textbf{return} $\Big( \bigwedge_{i \in [0, |\trace|]} \big( (\Time_i < \interval_{start} + \Time_0) \rightarrow \Pro(\trace^i, \TimeV^i, \varphi_1) \big) \Big) \land \Big( \bigvee_{j \in [0, |\trace|]} \big( \InInt{j} \land \Pro(\trace, \TimeV, \G_{[0, \Time_j - \Time_0)} \varphi_1) \land \Pro(\trace^j, \TimeV^j, \varphi_2) \big) \Big)$
		    \Else
		        \State \textbf{return} $\Big( \bigwedge_{i \in [0, |\trace|]} \big( (\Time_i < \interval_{start} + \Time_0) \rightarrow \Pro(\trace^i, \TimeV^i, \varphi_1) \big) \Big) \land \Big( \bigvee_{j \in [0, |\trace|]} \big( \InInt{j} \land \Pro(\trace, \TimeV, \G_{[0, \Time_j - \Time_0)} \varphi_1) \land \Pro(\trace^j, \TimeV^j, \varphi_2) \big) \lor \varphi_1 \U_{(\interval - (\Time_{|\trace|} - \Time_0)} \varphi_2 \Big)$
	        \EndIf
        \Else
            \State \textbf{return} $\big( \bigwedge_{i \in [0, |\trace|]} \Pro(\trace^i, \TimeV^i, \varphi_1) \big) \land \varphi_1 \U_{(\interval - (\Time_{|\trace|} - \Time_0)} \varphi_2 $
		\EndIf
		\EndFunction
	\end{algorithmic}
\end{algorithm*}

%% file: smt_prog_example.tex
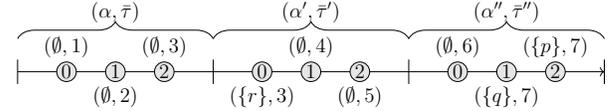
\begin{figure}
    \centering 
    \scalebox{0.65}{
        \begin{tikzpicture}
        
            
            \draw [->] (-0.5,0) -- (11.5,0);

            \draw (-0.5,-0.25) -- (-0.5,0.25);
            \draw (3.5,-0.25) -- (3.5,0.25);
            \draw (7.5,-0.25) -- (7.5,0.25);
            \draw (11.5,-0.25) -- (11.5,0.25);
            
            \large
            \draw (0.5,0) node[above, yshift=1.5mm]{{$(\emptyset, 1)$}};
            \draw [fill = black!10] (0.5,0) circle (0.2) node[]{{$0$}};
            \draw (1.5,0) node[below, yshift=-1.5mm]{ {$(\emptyset, 2)$}};
            \draw [fill = black!10] (1.5,0) circle (0.2) node[]{{$1$}};
            \draw (2.5,0) node[above, yshift=1.5mm]{ {$(\emptyset, 3)$}};
            \draw [fill = black!10] (2.5,0) circle (0.2) node[]{{$2$}};
            
            \draw (4.5,0) node[below, yshift=-1.5mm]{ {$(\{r\}, 3)$}};
            \draw [fill = black!10] (4.5,0) circle (0.2) node[]{ {$0$}};
            \draw (5.5,0) node[above, yshift=1.5mm]{ {$(\emptyset, 4)$}};
            \draw [fill = black!10] (5.5,0) circle (0.2) node[]{ {$1$}};
            \draw (6.5,0) node[below, yshift=-1.5mm]{ {$(\emptyset, 5)$}};
            \draw [fill = black!10] (6.5,0) circle (0.2) node[]{ {$2$}};
            
            \draw (8.5,0) node[above, yshift=1.5mm]{ {$(\emptyset, 6)$}};
            \draw [fill = black!10] (8.5,0) circle (0.2) node[]{ {$0$}};
            \draw (9.5,0) node[below, yshift=-1.5mm]{ {$(\{q\}, 7)$}};
            \draw [fill = black!10] (9.5,0) circle (0.2) node[]{ {$1$}};
            \draw (10.5,0) node[above, yshift=1.5mm]{ {$(\{p\}, 7)$}};
            \draw [fill = black!10] (10.5,0) circle (0.2) node[]{ {$2$}};
            
			\draw [decorate,decoration={brace,amplitude=10pt},xshift=0pt,yshift=5pt] (-0.5,0.5) -- (3.5,0.5) node 
			[black,midway,yshift=0.55cm] {$(\alpha, \TimeV)$};
			\draw [decorate,decoration={brace,amplitude=10pt},xshift=0pt,yshift=5pt] (3.5,0.5) -- (7.5,0.5) node 
			[black,midway,yshift=0.55cm] { $(\alpha', \TimeV')$};
			\draw [decorate,decoration={brace,amplitude=10pt},xshift=0pt,yshift=5pt] (7.5,0.5) -- (11.5,0.5) node 
			[black,midway,yshift=0.55cm] { $(\alpha'', \TimeV'')$};
        \end{tikzpicture}
    }
    \caption{Progression example.}
    \label{fig:progression}
\end{figure}

%% file: smt.tex
\ \\

\section{SMT-based Solution}
\label{sec:smt}

\subsection{SMT Entities}

SMT entities represent (1) sub-formulas of the \MTL specification, and (2) variables used to represent the distributed computation. 
After we have the verdicts for each of the individual sub-formulas, we use the progression laws discussed in Section~\ref{sec:progress} to construct the formula for the future computations. 


\vspace{2mm}
\noindent \textbf{Distributed Computation} We represent a distributed computation $(\Events, \hb)$ by function $f: \Events \rightarrow \{0, 1, \ldots, |\Events|-1\}$.
To represent the happen-before relation, we define a $\Events \times \Events$ matrix called $\hbSet$ where $\hbSet[e^i_\RTime][e^j_{\RTime'}] = 1$ represents $e^i_\RTime \hb e^j_{\RTime'}$ for $e^i_\RTime, e^j_{\RTime'} \in \Events$.
Also, if $|\RTime - \RTime'| \geq \epsilon$ then $\hbSet[e^i_\RTime][e^j_{\RTime'}] = 1$, else $\hbSet[e^i_\RTime][e^j_{\RTime'}] = 0$. This is all done in the pre-processing phase of the algorithm and in the rest of the paper, we represent events by the set $\Events$ and a happen-before relation by $\hb$ for simplicity.

In order to represent the possible time of occurrence of an event, we define a function $\delta: \Events \rightarrow \wholePlusSet$, where
$$\forall e^i_\RTime \in \Events. \exists \RTime' \in [\max\{0, \RTime - \epsilon + 1\}, \RTime + \epsilon - 1]. \delta(e^i_\RTime) = \RTime'$$
To connect events, $\Events$, and propositions, $\Pred$, on which the \MTL formula $\varphi$ is constructed, we define a boolean function $\mu : \Pred \times \Events \rightarrow \{ \tru, \fals \}$. For formulas involving non-boolean variables (e.g., $x_1 + x_2 \leq 7$), we can update the function $\mu$ accordingly. We represent a sequence of consistent cuts that start from $\{\}$ and end in $\Events$, we introduce an {\em uninterpreted function} $\rho: \wholePlusSet \rightarrow 2^\Events$ to reach a verdict given, it satisfies all the constrains explained in~\ref{subsec:smt_const}. Lastly, to represent the sequence of time associated with the sequence of consistent cuts, we introduce a function $\Time : \wholePlusSet \rightarrow \wholePlusSet$.

\subsection{SMT Constrains}
\label{subsec:smt_const}

Once we have the necessary SMT entities, we move onto including the constrains for both generating a sequence of consecutive cuts and also representing the \MTL formula as a SMT constrain.

\vspace{2mm}
\noindent \textbf{Consistent cut constrains over $\rho$:} In order to make sure the sequence of cuts represented by the uninterpreted function $\rho$, is a sequence of consistent cuts, i.e., they follow the happen-before relations between events in the distributed system:
$$\forall i \in [0, |\Events|]. \forall e, e' \in \Events. \Big( (e' \hb e) \land \big( e \in \rho(i) \big) \Big) \rightarrow \big( e' \in \rho(i) \big)$$
Next, we make sure that in the sequence of consistent cuts, the number of events present in a consistent cut is one more than the number of events that were present in the consistent cut before it:
$$\forall i \in [0, |\Events|). \mid\rho(i+1)\mid = \mid\rho(i)\mid + 1$$
Next, we make sure than in the sequence of consistent cuts, each consistent cut includes all the events that were present in the consistent cut before it, i.e, it is a subset of the consistent cut prior in the sequence.
$$\forall i \in [0, |\Events|]. \rho(i) \subset \rho(i+1)$$
The sequence of consistent cuts starts from $\{\}$ and ends at $\Events$.
$$\rho(0) = \emptyset; \; \rho(|\Events|) = \Events$$
The sequence of time reflects the time of occurrence of the event that has just been added to the sequence of consistent cut:
$$\forall i \geq 1. \Time(i) = \delta(e^i_\RTime) \text{, such that } \rho(i) - \rho(i-1) = \{e^i_\RTime\}$$
And finally, we make sure the monotonosity of time is maintained in the sequence of time 
$$\forall i \in [0, |\Events|). \Time(i + 1) \geq \Time(i)$$
\textbf{Constrains for \MTL formulas over $\rho$:} These constrains will make sure that $\rho$ will not only represent a valid sequence of consistent cuts but also make sure that the sequence of consistent cuts satisfy the \MTL formula. As is evident, a distributed computation can often yield two contradicting evaluation. Thus, we need to check for both satisfaction and violation for all the sub-formulas in the \MTL formula provided.
Note that monitoring any \MTL formula using our progression rules will result in monitoring sub-formulas which are atomic propositions, eventually and globally temporal operators. Below we mention the SMT constrain for each of the different sub-formula. Violation (resp. satisfaction) for atomic proposition and eventually (resp. globally) constrain will be the negation of the one mentioned.
\begin{align*}
    \varphi = \p & & \bigvee_{e \in \front(\rho(0))} \mu[\p, e] = \tru, \text{for } \p \in \Pred \\
    & &~~~~\text{(satisfaction, i.e., $\top$)}\\
    \varphi = \G_\interval \varphi & & \exists i \in [0, |\Events|]. \Time(i) - \Time(0) \in \interval \land \rho(i) \not\models \varphi \\
    & &~~~~\text{(violation, i.e., $\bot$)} \\
    \varphi = \F_\interval \varphi & & \exists i \in [0, |\Events|]. \Time(i) - \Time(0) \in \interval \land \rho(i) \models \varphi \\
    & &~~~~\text{(satisfaction, i.e., $\top$)}
\end{align*}
A satisfiable SMT instance denotes that the uninterpreted function was not only able to generate a valid sequence of consistent cuts but also that the sequence satisfies or violates the \MTL formula given the computation. This result is then fed to the progression cases to generate the final verdict.

\subsection{Segmentation and Parallelization of Distributed Computation}

We know that predicate detection, let alone runtime verification, is NP-complete~\cite{garg2002} in the size of the system (number of processes). This complexity grows to higher classes when working with nested temporal operators. To make the problem computationally viable, we aim to chop the computation, $(\Events, \hb)$ into $g$ segments, $(\seg_1, \hb), (\seg_2, \hb), \cdots, (\seg_g, \hb)$. This involves creating small SMT-instances for each of the segments which improves the runtime of the overall problem. In a computation of length $l$, if we were to chop it into $g$ segments, each segment would of the length $\frac{l}{g} + \epsilon$ and the set of events included in it can be given by:

\begin{align*}
\seg_j &= \Big\{ e^i_\RTime \mid \RTime \in \bigg[max\big(0, \frac{(j-1)\times l}{g} - \epsilon\big), \frac{j \times l}{g}\bigg] \land \\
&~~~ i \in [1, \mid\Proc\mid]\Big\}
\end{align*}

Note that monitoring of a segment should include the events that happened within $\epsilon$ time of the segment actually starting since it might include events that are concurrent with some other events in the system not accounted for in the previous segment.


%% file: eval.tex
\section{Case Study and Evaluation}
\label{sec:eval}

In this section, we analyze our SMT-based solution.
We note that we are not concerned about data collections, data transfer, etc, as given a distributed 
setting, the runtime of the actual SMT encoding will be the most dominating aspect of the monitoring 
process.
We evaluate our proposed solution using traces collected from benchmarks of the tool 
\code{UPPAAL}~\cite{lpy97}\footnote{\code{\footnotesize UPPAAL} is a model checker for a network 
of timed automata. The tool-set is accompanied by a set of benchmarks for real-time systems. Here, 
we assume that the components of the network are partially synchronized.} models 
(Section~\ref{sec:uppaal}) and a case study involving smart contracts over multiple blockchains 
(Section~\ref{sec:blockchain}).

\subsection{UPPAAL Benchmarks}
\label{sec:uppaal}

\subsubsection{Setup} We base our synthetic experiments on $3$ different \code{UPPAAL} benchmark models described in~\cite{uppaal04}. 
{\em The Train Gate} models a railway control system which controls access to a bridge. The bridge 
is controlled by a gate/operator and can be accessed by one train at a time. We monitor two 
properties:
%
\begin{align*}
    \varphi_1 &= (\bigwedge_{i \in \Proc} \neg \texttt{Train[i].Cross}) \;\U\; \texttt{Train[1].Cross} \\
    \varphi_2 &= \bigwedge_{i \in \Proc} \G \big( \texttt{Train[i].Appr} \rightarrow \\
    & ~~~\F (\texttt{Gate.Occ} \;\U\; \texttt{Train[i].Cross}) \big)
\end{align*}
where $\Proc$ is the set of trains.

{\em Fischer’s Protocol} is a mutual exclusion protocol for $n$ processes. We verify first, that no two 
process (\texttt{P}) enter the critical section (\texttt{cs}) at the same time and second, all request 
(\texttt{req}) should be followed by the processes able to access the critical section within some 
time. 
\begin{align*}
    \varphi_3 &= \G (\sum_{i \in \Proc} \texttt{P[i].cs} \leq 1) \\
    \varphi_4 &= \G (\bigwedge_{i \in \Proc} \texttt{P[i].req} \rightarrow \F_\interval \texttt{P[i].cs} ) 
\end{align*}
{\em The Gossiping People} is a model consisting of $n$ people who wish to share their secret with each other. We monitor first, that each \texttt{Person} gets to know about everyone else's \texttt{secret} within some time bound and second, each \texttt{Person} has \texttt{secrets} to share infinitely often.
\begin{align*}
    \varphi_5 &= \F_\interval (\bigwedge_{i, j \in \Proc} (i \neq j) \rightarrow \texttt{Person[i].secret[j]} ) \\
    \varphi_6 &= \bigwedge_{i \in \Proc} \G (\F_\interval \texttt{Person[i].secrets})
\end{align*}
%

Each experiment involves two steps: (1) distributed computation/trace generation and (2) trace 
verification. For each \code{UPPAAL} model, we consider each pair of consecutive events are $0.1 
s$ apart, i.e., there are $10$ events per second per process.
For our verification step, our monitoring algorithm executes on the generated computation and verifies it against an \MTL specification.
We consider the following parameters (1) primary which includes time synchronization constant ($\epsilon$), (2) \MTL formula under monitoring, (3) number of segments ($g$), (3) computation length ($l$), (4) number of processes in the system ($\Proc$), and (5) event rate. We study the runtime of our monitoring algorithm against each of these parameters.
We use a machine with $2x$ Intel Xeon Platinum $8180$ ($2.5$ Ghz) processor, $768$ GB of RAM, $112$ vcores with gcc version $9.3.1$.

\subsubsection{Analysis} We now study each of the parameters individually and analyze how it effects the runtime of our monitoring approach. All results correspond to $\epsilon = 15 ms$, $|\Proc| = 2$, $g = 15$, $l = 2sec$, a event rate of $10 events/sec$ and $\varphi_4$ as the specification unless mentioned otherwise.

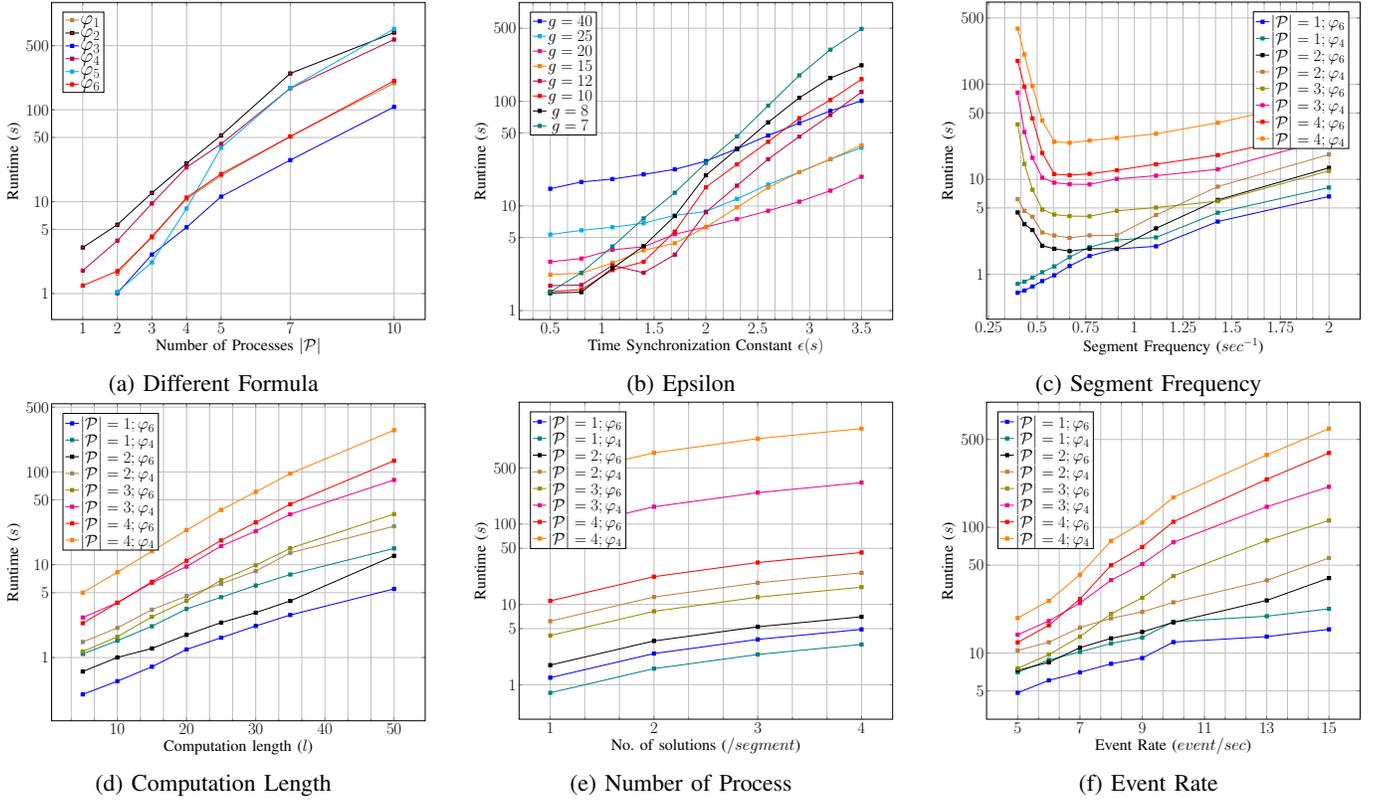
\begin{figure*}[t]
  \centering
  \subcaptionbox{Different Formula\label{graph:diffFormula}}
      {\scalebox{0.3}{\input{graph_diffFormula}}}
    \hfill
    \subcaptionbox{Epsilon\label{graph:epsilon}}
      {\scalebox{0.3}{\input{graph_epsilon}}}
    \hfill
    \subcaptionbox{Segment Frequency\label{graph:segFreq}}
      {\scalebox{0.3}{\input{graph_segFreq}}}
    \hfill
    \subcaptionbox{Computation Length\label{graph:compLength}}
      {\scalebox{0.3}{\input{graph_compLength}}}
    \hfill
    \subcaptionbox{Number of Process\label{graph:numProcess}}
      {\scalebox{0.3}{\input{graph_numProcess}}}
    \hfill
    \subcaptionbox{Event Rate\label{graph:eventRate}}
      {\scalebox{0.3}{\input{graph_eventRate}}}
      
    \caption{Impact of different parameters on synthetic data}

 \vspace{-2mm}
\end{figure*}

\vspace{2mm}
\noindent {\bf Impact of different formula.} 
%
Fig.~\ref{graph:diffFormula} shows that runtime of the monitor depends on two factors: 
the number of sub-formulas and the depth of nested temporal operators. Comparing 
$\varphi_3$ and $\varphi_6$, both of which consists of the same number of predicates but since 
$\varphi_6$ has recursive temporal operators, it takes more time to verify and the runtime is 
comparable to $\varphi_1$, which consists of two sub-formulas. This is because verification of the 
inner temporal formula often requires observing states in the next segment in order to come to the 
final verdict. This accounts for the more runtime for the monitor.

\vspace{2mm}
\noindent {\bf Impact of epsilon.} Increasing the value of time synchronization constant ($\epsilon$), 
increases the possible number of concurrent events that needs to be considered. This increases the 
complexity of verifying the computation and there-by increasing the runtime of the algorithm. In 
addition to this, higher values of $\epsilon$ also correspond to more number of possible traces that 
are possible and should be taken into consideration. We observe that the runtime increases 
exponentially with increasing the time synchronization constant in Fig.~\ref{graph:epsilon}. An 
interesting observation is with longer segment length, the runtime increases at a higher rate than with
shorter segment length. This is because with longer segment length and higher $\epsilon$, it equates 
to a larger number of possible traces that the monitoring algorithm needs to take into consideration. 
This increases the overall runtime of the verification algorithm by a considerable amount and at a 
higher pace.

      


\vspace{2mm}
\noindent {\bf Impact of segment frequency.} Increasing the segment frequency makes the length of 
each segment lower and thus verifying each segment involves consideration of a lower number of 
events. We observe the effect of segment frequency on the runtime of our verification algorithm in 
Fig.~\ref{graph:segFreq}. With increasing the segment frequency, the runtime decreases unless it 
reaches a certain value (here it is $\approx 0.6$) after which the benefit of working with a lower 
number of events is overcast by the time required to setup each SMT instances. Working with 
higher number of segments equates to solving more number of SMT problem for the same 
computation length. Setting up the SMT problem requires a considerable amount of time which is 
seen by the slight increase in runtime for higher values of segment frequency.

\vspace{2mm}
\noindent {\bf Impact of computation length.} As it can be inferred from the previous results, the 
runtime of our verification algorithm is majorly dictated by the number of events in the computation. 
Thus, when working with a longer computation, keeping the maximum clock skew and the number of 
segments constant, we should see a longer verification time as well. Results in 
Fig.~\ref{graph:compLength} makes the above claim true.

\vspace{2mm}
\noindent {\bf Impact of number truth values per segment.} In order to take into consideration all 
possible truth values of a computation, we execute the SMT problem multiple times, with the verdict 
of all previous executions being added to the SMT problem such that no two verdict is repeated. Here in 
Fig.~\ref{graph:numProcess} we see that the runtime is linearly effected by increasing number of 
distinct verdicts. This is because, the complexity of the problem that the SMT is trying to solve does 
not change when trying to evaluate to a different solution.

\vspace{2mm}
\noindent {\bf Impact of event-rate.} Increasing the event rate involves more number of events that 
needs to be processes by our verification algorithm per segment and thereby increasing the runtime 
at an exponential rate as seen in Fig.~\ref{graph:eventRate}. We also observe that with higher 
number of processes, the rate at which the runtime of our algorithm increases is higher for the same 
increase in event rate.

\subsection{Blockchain}
\label{sec:blockchain}

\subsubsection{Setup}

We implemented the following cross-chain protocols from \cite{xue2021hedging}: two-party swap, 
multi-party swap, and auction. The protocols were written as smart contracts in \code{Solidity} and 
tested using \code{Ganache}, a tool that creates mocked \code{Ethereum} blockchains. Using a 
single mocked chain, we mimicked cross-chain protocols via several (discrete) tokens and smart 
contracts, which do not communicate with each other.

We use the hedged two-party swap example  from \cite{xue2021hedging} to describe our 
experiments. The implementation of the other two protocols are similar. Suppose Alice would like to 
exchange her apricot tokens with Bob's banana tokens, using the hedged two-party swap protocol 
shown in Fig.~\ref{fig:intro1}.  This protocol provides protection for parties compared to a standard 
two-party swap protocol \cite{tiernolan}, in that if one party locks their assets to exchange which is 
refunded later, this party gets a premium as compensation for locking their assets. The protocol 
consists of six steps to be executed by Alice and Bob in turn. In our example, we let the amount of 
tokens they are exchanging be 100 ERC20 tokens and the premium $p_b$ be 1 token and 
$p_a+p_b$ be 2 tokens. We deploy two contracts on both apricot blockchain(the contract is 
denoted as $ApricotSwap$) and banana blockchain (denoted as $BananaSwap$) by mimicking the 
two blockchains on Ethereum. Denote the time that they reach an agreement of the swap as 
$startTime$. $\Delta$ is the maximum time for parties to observe the state change of contracts by 
others and take a step to make changes on contracts. In our experiment, $\Delta = 500$ 
milliseconds. By the definition of the protocol, the execution should be:
\begin{itemize}
    \item Step 1. Alice deposits 2 tokens as premium in  $BananaSwap$ before $\Delta$ elapses after $startTime$ .
    \item Step 2. Bob should deposit 1 token as premium in $ApricotSwap$ before $2\Delta$ elapses after $startTime$.
    \item Step 3. Alice escrows her 100 ERC20 tokens to $ApricotSwap$ before $3\Delta$ elapses after $startTime$.
    \item Step 4. Bob escrows her 100 ERC20 tokens to $BananaSwap$ before $4\Delta$ elapses after $startTime$.
    \item Step 5. Alice sends the preimage of the hashlock to $BananaSwap$ to redeem Bob's 100 tokens before $5\Delta$ elapses after $startTime$. Premium is refunded.
    \item Step 6. Bob sends the preimage of the hashlock to $ApricotSwap$  to redeem Alice's 100 tokens before $6\Delta$ elapses after $startTime$. Premium is refunded.
\end{itemize}   

If all parties all conforming, the protocol is executed as above. Otherwise, some asset refund and premium redeem events will be triggered to resolve the case where some party deviates. To avoid distraction, we do not provide details here.

Each smart contract provides functions to let parties deposit premiums \texttt{DepositPremium()}, escrow an asset \texttt{EscrowAsset()}, send a secret to redeem assets \texttt{RedeemAsset()}, refund the asset if it is not redeemed after timeout, \texttt{RefundAsset()}, and counterparts for premiums \texttt{RedeemPremium()} and \texttt{RefundPremium()}. Whenever a function is called successfully (meaning the transaction sent to the blockchain is included in a block), the blockchain emits an event that we then capture and log. The event interface is provided by the Solidity language. For example, when a party successfully calls \texttt{DepositPremium()}, the \texttt{PremiumDeposited} event emits on the blockchain. We then capture and log this event, allowing us to view the values of \texttt{PremiumDeposited}'s declared fields: the time when it emits, the party that called \texttt{DepositPremium()}, and the amount of premium sent. Those values are later used in the monitor to check against the specification.

\subsubsection{Log Generation and Monitoring}

Our tests simulated different executions of the protocols and generated 1024, 4096, and 3888 different sets of logs for the aforementioned protocols, respectively. We use the hedged two-party swap as an example to show how we generate different logs to simulate different execution of the protocol. On each contract, we enforce the order of those steps to be executed. For example, step 3 \texttt{EscrowAsset()} on the $ApricotSwap$ cannot be executed before Step 1 is taken, i.e. the premium is deposited. This enforcement in the contract restricted the number of possible different states in the contract. Assume we use a binary indicator to denote whether a step is attempted by the corresponding party. $1$ denotes a step is attempted, and $0$ denotes this step is skipped. If the previous step is skipped, then the later step does not need to be attempted since it will be rejected by the contract. We use an array to denote whether each step in taken for each contract. On each contract, the different execution of those steps can be [1,1,1] means all steps are attempted, or [1,1,0] meaning the last step is skipped, and so on. Each chain has 4 different executions. We take the Cartesian product of arrays of two contracts to simulate different combinations of executions on two contracts. Furthermore, if a step is attempted, we also simulate whether the step is taken late, or in time. Thus we have $2^6$ possibilities of those 6 steps. In summary, we succeeded generating $4 \cdot 4 \cdot 2^6=1024$ different logs.

In our testing, after deploying the two contracts, we iterate over a 2D array of size $1024 \times 12$, and each time takes one possible execution denoted as an array length of 12 to simulate the behavior of participants. For example, $[1,0,1,1,1,1,1,1,1,1,1,1]$ means the first step is attempted however it is late, and the steps after second step
are all attempted in time. Indexed from 0, the even index denotes if a step is attempted or not and the odd index denotes the former step is attempted in time or late. By the indicator given by the array, we let parties attempt to call a function of the contract or just skip. In this way, we produce $1024$ different logs containing the events emitted in each iteration.

We check the policies mentioned in~\cite{xue2021hedging}: liveness, safety, and ability to hedge 
against sore loser attacks. {\em Liveness} means that Alice should deposit her premium on the 
banana blockchain within $\Delta$ from when the swap started($ \F_{[0, \Delta)} 
\texttt{ban.premium\_deposited(alice)}$) and then Bob should deposit his premiums, and then they 
escrow their assets to exchange, redeem their assets (i.e. the assets are swapped), and the 
premiums are refunded. In our testing, we always call a function to settle all assets in the contract if 
the asset transfer is triggered by timeout. Thus, in the specification, we also check all assets are 
settled:
\begin{align*}
\varphi_{\liveness} &= \F_{[0, \Delta)} \texttt{ban.premium\_deposited(alice)} \land \\
& \F_{[0, 2\Delta)} \texttt{apr.premium\_deposited(bob)} \land \\
& \F_{[0, 3\Delta)} \texttt{apr.asset\_escrowed(alice)} \land \\
& \F_{[0, 4\Delta)} \texttt{ban.asset\_escrowed(bob)} \land \\
& \F_{[0, 5\Delta)} \texttt{ban.asset\_redeemed(alice)} \land \\
& \F_{[0, 6\Delta)} \texttt{apr.asset\_redeemed(bob)} \land \\
& \F_{[0, 5\Delta)} \texttt{ban.premium\_refunded(alice)} \land \\
& \F_{[0, 6\Delta)} \texttt{apr.premium\_refunded(bob)} \land \\
& \F_{[6\Delta, \infty)} \texttt{apr.all\_asset\_settled(any)} \land \\
& \F_{[5\Delta, \infty)} \texttt{ban.all\_asset\_settled(any)}
\end{align*}
{\em Safety} is provided only for conforming parties, since if one party is deviating and behaving 
unreasonably, it is out of the scope of the protocol to protect them. Alice should always deposit her 
premium first to start the execution of the protocol($\F_{[0, \Delta)} 
\texttt{ban.premium\_deposited(alice)}$) and proceed if Bob proceeds with the next step. For 
example, if Bob deposits his premium, then Alice should always go ahead and escrow her asset to 
exchange($ \F_{[0, 2\Delta)} \texttt{apr.premium\_deposited(bob)} \rightarrow \F_{[0, 3\Delta)} 
\texttt{apr.asset\_escrowed(alice)}$). Alice should never release her secret if she does not redeem, 
which means Bob should not be able to redeem unless Alice redeems, which is expressed as $\neg 
\texttt{apr.asset\_redeemed(bob)} \U$ $\texttt{ban.asset\_redeemed(alice)}$:

{\small 
\begin{align*}
\varphi_{\mathsf{alice\_conform}} &= \F_{[0, \Delta)} \texttt{ban.premium\_deposited(alice)} \land \\
& \big( \F_{[0, 2\Delta)} \texttt{apr.premium\_deposited(bob)} \rightarrow \\
&~~~\F_{[0, 3\Delta)} \texttt{apr.asset\_escrowed(alice)} \big) \, \land \\
& \big( \F_{[0, 4\Delta)} \texttt{ban.asset\_escrowed(bob)} \rightarrow \\
&~~~\F_{[0, 5\Delta)} \texttt{ban.asset\_redeemed(alice)} \big) \, \land \\
& \big( \neg \texttt{apr.asset\_redeemed(bob)} \U \\
&~~~\texttt{ban.asset\_redeemed(alice)} \big)
\end{align*}
}
By definition, safety means a conforming party does not end up with a negative payoff.  We track the 
assets transferred from parties and transferred to parties in our logs.
%
Thus, a conforming party is safe. e.g. Alice, is specified as the $\varphi_{alice\_safety}$:
\begin{align*}
\varphi_{\mathsf{alice\_safety}} =& \varphi_{\mathsf{alice\_conform}} \rightarrow \\
\big( \sum_{\texttt{TransTo = alice}} \texttt{amount} &\geq \sum_{\texttt{TransFrom = alice}} \texttt{amount} \big)
\end{align*}
To enable a conforming party to hedge against the sore loser attack if they escrow assets to 
exchange which is refunded in the end, our protocol should guarantee the aforementioned party get 
a premium as compensation, which is expressed as $\varphi_{\mathsf{alice\_hedged}}$:
\begin{align*}
\varphi_{\mathsf{alice\_hedged}}= &\F \big( \varphi_{\mathsf{alice\_conform}} \land \\
& \texttt{apr.asset\_escrowed(alice)} \land \\
& \texttt{apr.asset\_refunded(any)} \big) \rightarrow \\
& \F \big( \sum_{\texttt{TransferTo = alice}} \texttt{amount} \geq \\
& \sum_{\texttt{TransferFrom = alice}} \texttt{amount}\\
&+\texttt{apr.premium.amount} \big)
\end{align*}


\subsubsection{Analysis of Results}

\input{graph_blockc}

We put our monitor to test the traces generated by the Truffle-Ganache framework. To monitor the 
2-party swap protocol we do not divide the trace into multiple segments due to the low number of 
events that are involved in the protocol. On the other hand both, 3-party swap and auction protocol 
involves a higher number of events and thus we divide the trace into two segments ($g=2$). In 
Fig.~\ref{graph:blockc}, we show how the runtime of the monitor is effected by the number of events 
in each transaction log.

Additionally, we generate transaction logs with different values for deadline ($\Delta$) and time 
synchronization constant ($\epsilon$) to put the safety of the protocol in jeopardy. We observe both 
$\tru$ and $\fals$ verdict when $\epsilon \gtrapprox \Delta$. This is due to the non deterministic 
time stamp owning to the assumption of a partially synchronous system. The observed time stamp 
of each event can at most be off by $\epsilon$. Thus, we recommend not to use a value of $\Delta$ 
that is comparable to the value of $\epsilon$ when designing the smart contract.

%% file: graph_diffFormula.tex
        \pgfplotsset{every tick label/.append style={font=\huge}}
    
    \pgfplotsset{every axis/.append style={
                       label style={font=\huge}}}
        \begin{tikzpicture}
            \begin{axis}[
                width=\textwidth,
                grid=both,
                minor tick num=2,
                xlabel={Number of Processes $|\Proc|$},
                ylabel={Runtime ($s$)},
                legend style={nodes={scale=1, transform shape}, font=\Huge},
            y filter/.code=\pgfmathparse{#1 * 2.30258509299},
            ymode = log,
            log ticks with fixed point,
            ytick = {1, 5, 10, 50, 100, 500},
            y label style={font=\huge},
            x label style={font=\huge},
            xtick = {1, 2, 3, 4, 5, 7, 10},
            legend pos = north west
                ]
                \addplot[color=brown, mark=square*, ultra thin, mark options={solid, fill=brown, scale=1}] coordinates {
(2, 0.2195845262)
(3, 0.6259294927)
(4, 1.02820512)
(5, 1.283233365)
(7, 1.711714457)
(10, 2.289626851)
                };
                \addlegendentry{$\varphi_1$}
                
                \addplot[color=black, mark=square*, ultra thin, mark options={solid, fill=red, scale=01}] coordinates {
(1, 0.5013331786)
(2, 0.7493497606)
(3, 1.096110175)
(4, 1.416290934)
(5, 1.721332381)
(7, 2.395967313)
(10, 2.84078086)
                };
                \addlegendentry{$\varphi_2$}
                
                \addplot[color=blue, mark=square*, ultra thin, mark options={solid, fill=blue, scale=1}] coordinates {
(2, 0.004751155591)
(3, 0.4248816366)
(4, 0.7209857442)
(5, 1.055378331)
(7, 1.45272154)
(10, 2.031153873)
                };
                \addlegendentry{$\varphi_3$}
                
                \addplot[color=purple, mark=square*, ultra thin, mark options={solid, fill=purple, scale=1}] coordinates {
(1, 0.2494429614)
(2, 0.5761108941)
(3, 0.9801852369)
(4, 1.373518855)
(5, 1.628583042)
(7, 2.234213475)
(10, 2.765038632)
                };
                \addlegendentry{$\varphi_4$}
                
                \addplot[color=cyan, mark=square*, ultra thin, mark options={solid, fill=cyan, scale=1}] coordinates {
(2, 0.01786771896)
(3, 0.3400473177)
(4, 0.924692703)
(5, 1.587508921)
(7, 2.237159036)
(10, 2.880948431)
                };
                \addlegendentry{$\varphi_5$}

                \addplot[color=red, mark=square*, ultra thin, mark options={solid, fill=red, scale=1}] coordinates {
(1, 0.08707120591)
(2, 0.244029589)
(3, 0.612995656)
(4, 1.044186851)
(5, 1.301333895)
(7, 1.708335903)
(10, 2.315605133)
                };
                \addlegendentry{$\varphi_6$}

            \end{axis}
        \end{tikzpicture}

%% file: graph_epsilon.tex
        \pgfplotsset{every tick label/.append style={font=\huge}}
    
    \pgfplotsset{every axis/.append style={
                       label style={font=\huge}}}
        \begin{tikzpicture}
            \begin{axis}[
                width=\textwidth,
                grid=both,
                minor tick num=2,
                xlabel={Time Synchronization Constant $\epsilon (s)$},
                ylabel={Runtime ($s$)},
                legend style={nodes={scale=1, transform shape}, font=\huge},
            y filter/.code=\pgfmathparse{#1 * 2.30258509299},
            ymode = log,
            log ticks with fixed point,
            ytick = {1, 5, 10, 50, 100, 500},
            y label style={font=\huge},
            x label style={font=\huge},
            xtick = {0, 0.5, 1, 1.5, 2, 2.5, 3, 3.5},
            legend pos = north west
                ]
                \addplot[color=blue, mark=square*, ultra thin, mark options={solid, fill=blue, scale=1}] coordinates {
(0.5,1.164602652)
(0.8,1.229548744)
(1.1,1.257236858)
(1.4,1.30165493)
(1.7,1.349533948)
(2,1.428769567)
(2.3,1.546394583)
(2.6,1.672937913)
(2.9,1.792391689)
(3.2,1.908485019)
(3.5,2.004321374)
                };
                \addlegendentry{$g = 40$}
                
                \addplot[color=cyan, mark=square*, ultra thin, mark options={solid, fill=cyan, scale=1}] coordinates {
(0.5,0.7273134774)
(0.8,0.7670073639)
(1.1,0.7966575766)
(1.4,0.8355891186)
(1.7,0.909876818)
(2,0.9465505106)
(2.3,1.067977945)
(2.6,1.204119983)
(2.9,1.322219295)
(3.2,1.447158031)
(3.5,1.556302501)
                };
                \addlegendentry{$g = 25$}
                
                \addplot[color=magenta, mark=square*, ultra thin, mark options={solid, fill=magenta, scale=1}] coordinates {
(0.5,0.4665710724)
(0.8,0.4980347237)
(1.1,0.5809249757)
(1.4,0.6098077693)
(1.7,0.7295697263)
(2,0.7989957344)
(2.3,0.8733206018)
(2.6,0.9542425094)
(2.9,1.041392685)
(3.2,1.146128036)
(3.5,1.278753601)
                };
                \addlegendentry{$g = 20$}
                
                \addplot[color=orange, mark=square*, ultra thin, mark options={solid, fill=orange, scale=1}] coordinates {
(0.5,0.3440777385)
(0.8,0.3574774541)
(1.1,0.4555148327)
(1.4,0.5761108941)
(1.7,0.6438078631)
(2,0.7971844143)
(2.3,0.986018902)
(2.6,1.176091259)
(2.9,1.322219295)
(3.2,1.447158031)
(3.5,1.579783597)
                };
                \addlegendentry{$g = 15$}
                
                \addplot[color=purple, mark=square*, ultra thin, mark options={solid, fill=purple, scale=1}] coordinates {
(0.5,0.2382468865)
(0.8,0.2443266102)
(1.1,0.4321351578)
(1.4,0.3615767508)
(1.7,0.5341276835)
(2,0.9403969379)
(2.3,1.19384782)
(2.6,1.447158031)
(2.9,1.662757832)
(3.2,1.86923172)
(3.5,2.089905111)
                };
                \addlegendentry{$g = 12$}
                
                \addplot[color=red, mark=square*, ultra thin, mark options={solid, fill=red, scale=1}] coordinates {
(0.5,0.1817292849)
(0.8,0.2006316548)
(1.1,0.3872474114)
(1.4,0.4667490255)
(1.7,0.7543177505)
(2,1.179321945)
(2.3,1.397905264)
(2.6,1.612783857)
(2.9,1.838849091)
(3.2,2.012837225)
(3.5,2.212187604)
                };
                \addlegendentry{$g = 10$}
                
                \addplot[color=black, mark=square*, ultra thin, mark options={solid, fill=black, scale=1}] coordinates {
(0.5,0.1669626639)
(0.8,0.176322821)
(1.1,0.4050046651)
(1.4,0.6161603128)
(1.7,0.90330708)
(2,1.293008787)
(2.3,1.544068044)
(2.6,1.799340549)
(2.9,2.033423755)
(3.2,2.222716471)
(3.5,2.344392274)
                };
                \addlegendentry{$g = 8$}
                
                \addplot[color=teal, mark=square*, ultra thin, mark options={solid, fill=teal, scale=1}] coordinates {
(0.5,0.1789769473)
(0.8,0.361727836)
(1.1,0.6127838567)
(1.4,0.8808135923)
(1.7,1.127104798)
(2,1.411619706)
(2.3,1.665580991)
(2.6,1.959041392)
(2.9,2.247973266)
(3.2,2.494154594)
(3.5,2.691965103)
                };
                \addlegendentry{$g = 7$}

            \end{axis}
        \end{tikzpicture}

%% file: graph_segFreq.tex
        \pgfplotsset{every tick label/.append style={font=\huge}}
    
    \pgfplotsset{every axis/.append style={
                       label style={font=\huge}}}
        \begin{tikzpicture}
            \begin{axis}[
                width=\textwidth,
                grid=both,
                minor tick num=2,
                xlabel={Segment Frequency ($sec^{-1}$)},
                ylabel={Runtime ($s$)},
                legend style={nodes={scale=1, transform shape}, font=\huge},
            y filter/.code=\pgfmathparse{#1 * 2.30258509299},
            x filter/.code=\pgfmathparse{1 / #1},
            ymode = log,
            log ticks with fixed point,
            ytick = {1, 5, 10, 50, 100, 500},
            y label style={font=\huge},
            x label style={font=\huge},
            xtick = {0, 0.25, 0.5, 0.75, 1, 1.25, 1.5, 1.75, 2},
            legend pos = north east
                ]
                \addplot[color=blue, mark=square*, ultra thin, mark options={solid, fill=blue, scale=1}] coordinates {
(0.5, 0.8197174804)
(0.7, 0.5557630388)
(0.9, 0.2952625398)
(1.1, 0.2674963219)
(1.3, 0.1929430791)
(1.5, 0.08707120591)
(1.7, -0.01029752854)
(1.9, -0.07100254026)
(2.1, -0.1282175949)
(2.3, -0.1707879327)
(2.5, -0.1943497492)
                };
                \addlegendentry{$|\Proc| = 1; \varphi_6$}
                
                \addplot[color=teal, mark=square*, ultra thin, mark options={solid, fill=teal, scale=1}] coordinates {
(0.5, 0.9132308711)
(0.7, 0.6492764295)
(0.9, 0.3887759304)
(1.1, 0.3610097126)
(1.3, 0.2864564697)
(1.5, 0.1805845966)
(1.7, 0.08321586216)
(1.9, 0.02251085043)
(2.1, -0.03470420419)
(2.3, -0.07727454201)
(2.5, -0.1008363585)
                };
                \addlegendentry{$|\Proc| = 1; \varphi_4$}
                
                \addplot[color=black, mark=square*, ultra thin, mark options={solid, fill=black, scale=1}] coordinates {
(0.5, 1.121180875)
(0.7, 0.7829941106)
(0.9, 0.4837649239)
(1.1, 0.2707086887)
(1.3, 0.2699996363)
(1.5, 0.244029589)
(1.7, 0.2690185617)
(1.9, 0.3009394753)
(2.1, 0.4665092681)
(2.3, 0.5275265876)
(2.5, 0.6511700733)
                };
                \addlegendentry{$|\Proc| = 2; \varphi_6$}
                
                \addplot[color=brown, mark=square*, ultra thin, mark options={solid, fill=brown, scale=1}] coordinates {
(0.5, 1.261181952)
(0.7, 0.9229951868)
(0.9, 0.6237660001)
(1.1, 0.4107097649)
(1.3, 0.4100007125)
(1.5, 0.3840306652)
(1.7, 0.4090196379)
(1.9, 0.4409405515)
(2.1, 0.6065103443)
(2.3, 0.6675276638)
(2.5, 0.7911711496)
                };
                \addlegendentry{$|\Proc| = 2; \varphi_4$}
                
                \addplot[color=olive, mark=square*, ultra thin, mark options={solid, fill=olive, scale=1}] coordinates {
(0.5, 1.089405995)
(0.7, 0.7702888877)
(0.9, 0.7027700934)
(1.1, 0.6689999089)
(1.3, 0.6106334097)
(1.5, 0.612995656)
(1.7, 0.6292658563)
(1.9, 0.6817012162)
(2.1, 0.8905116989)
(2.3, 1.162360176)
(2.5, 1.578879986)
                };
                \addlegendentry{$|\Proc| = 3; \varphi_6$}
                
                \addplot[color=magenta, mark=square*, ultra thin, mark options={solid, fill=magenta, scale=1}] coordinates {
(0.5, 1.425097098)
(0.7, 1.10597999)
(0.9, 1.038461196)
(1.1, 1.004691012)
(1.3, 0.9463245126)
(1.5, 0.9486867589)
(1.7, 0.9649569591)
(1.9, 1.017392319)
(2.1, 1.226202802)
(2.3, 1.498051278)
(2.5, 1.914571089)
                };
                \addlegendentry{$|\Proc| = 3; \varphi_4$}
                
                \addplot[color=red, mark=square*, ultra thin, mark options={solid, fill=red, scale=1}] coordinates {
(0.5, 1.518614538)
(0.7, 1.255092863)
(0.9, 1.159261938)
(1.1, 1.095711499)
(1.3, 1.058101995)
(1.5, 1.044186851)
(1.7, 1.055234596)
(1.9, 1.277842515)
(2.1, 1.641330788)
(2.3, 1.973815676)
(2.5, 2.247028333)
                };
                \addlegendentry{$|\Proc| = 4; \varphi_6$}
                
                \addplot[color=orange, mark=square*, ultra thin, mark options={solid, fill=orange, scale=1}] coordinates {
(0.5, 1.858935219)
(0.7, 1.595413544)
(0.9, 1.479582619)
(1.1, 1.43603218)
(1.3, 1.408422676)
(1.5, 1.384507532)
(1.7, 1.395555277)
(1.9, 1.618163196)
(2.1, 1.981651469)
(2.3, 2.314136357)
(2.5, 2.587349014)
                };
                \addlegendentry{$|\Proc| = 4; \varphi_4$}

            \end{axis}
        \end{tikzpicture}

%% file: graph_compLength.tex
        \pgfplotsset{every tick label/.append style={font=\huge}}
    
    \pgfplotsset{every axis/.append style={
                       label style={font=\huge}}}
        \begin{tikzpicture}
            \begin{axis}[
                width=\textwidth,
                grid=both,
                minor tick num=2,
                xlabel={Computation length ($l$)},
                ylabel={Runtime ($s$)},
                legend style={nodes={scale=1, transform shape}, font=\huge},
            y filter/.code=\pgfmathparse{#1 * 2.30258509299},
            ymode = log,
            log ticks with fixed point,
            ytick = {1, 5, 10, 50, 100, 500},
            y label style={font=\huge},
            x label style={font=\huge},
            xtick = {0, 10, 20, 30, 40, 50},
            legend pos = north west
                ]
                \addplot[color=blue, mark=square*, ultra thin, mark options={solid, fill=blue, scale=1}] coordinates {
(5, -0.3970257638)
(10, -0.2550186105)
(15, -0.09917975953)
(20, 0.08707120591)
(25, 0.2135939655)
(30, 0.3400948269)
(35, 0.4585495791)
(50, 0.7395720601)
                };
                \addlegendentry{$|\Proc| = 1; \varphi_6$}
                
                \addplot[color=teal, mark=square*, ultra thin, mark options={solid, fill=teal, scale=1}] coordinates {
(5, 0.03949343513)
(10, 0.1815005885)
(15, 0.3373394394)
(20, 0.5235904048)
(25, 0.6501131644)
(30, 0.7766140258)
(35, 0.895068778)
(50, 1.176091259)
                };
                \addlegendentry{$|\Proc| = 1; \varphi_4$}
                
                \addplot[color=black, mark=square*, ultra thin, mark options={solid, fill=black, scale=1}] coordinates {
(5, -0.1507818931)
(10, 0.001027218675)
(15, 0.09805376047)
(20, 0.244029589)
(25, 0.376958014)
(30, 0.4833067206)
(35, 0.6101470505)
(50, 1.096519739)
                };
                \addlegendentry{$|\Proc| = 2; \varphi_6$}
                
                \addplot[color=brown, mark=square*, ultra thin, mark options={solid, fill=cyan, scale=1}] coordinates {
(5, 0.1676717162)
(10, 0.3194808281)
(15, 0.5165073699)
(20, 0.6624831984)
(25, 0.7954116233)
(30, 0.93176033)
(35, 1.1286006598)
(50, 1.414973348)
                };
                \addlegendentry{$|\Proc| = 2; \varphi_4$}
                
                \addplot[color=olive, mark=square*, ultra thin, mark options={solid, fill=olive, scale=1}] coordinates {
(5, 0.06417418324)
(10, 0.2238750261)
(15, 0.4389903931)
(20, 0.612995656)
(25, 0.8339343163)
(30, 0.9945382551)
(35, 1.176878463)
(50, 1.546624271)
                };
                \addlegendentry{$|\Proc| = 3; \varphi_6$}
                
                \addplot[color=magenta, mark=square*, ultra thin, mark options={solid, fill=magenta, scale=1}] coordinates {
(5, 0.4313637642)
(10, 0.591064607)
(15, 0.806179974)
(20, 0.9801852369)
(25, 1.201123897)
(30, 1.361727836)
(35, 1.544068044)
(50, 1.913813852)
                };
                \addlegendentry{$|\Proc| = 3; \varphi_4$}
                
                \addplot[color=red, mark=square*, ultra thin, mark options={solid, fill=red, scale=1}] coordinates {
(5, 0.3705057217)
(10, 0.5897460882)
(15, 0.8167960315)
(20, 1.044186851)
(25, 1.261732603)
(30, 1.455997831)
(35, 1.652939229)
(50, 2.120917104)
                };
                \addlegendentry{$|\Proc| = 4; \varphi_6$}
                
                \addplot[color=orange, mark=square*, ultra thin, mark options={solid, fill=orange, scale=1}] coordinates {
(5, 0.6998377259)
(10, 0.9190780924)
(15, 1.146128036)
(20, 1.373518855)
(25, 1.591064607)
(30, 1.785329835)
(35, 1.982271233)
(50, 2.450249108)
                };
                \addlegendentry{$|\Proc| = 4; \varphi_4$}

            \end{axis}
        \end{tikzpicture}

%% file: graph_numProcess.tex
    \pgfplotsset{every tick label/.append style={font=\huge}}
    \pgfplotsset{every axis/.append style={
                       label style={font=\huge}}}
        \begin{tikzpicture}
            \begin{axis}[
                width=\textwidth,
                grid=both,
                minor tick num=2,
                xlabel={No. of solutions ($/segment$)},
                ylabel={Runtime ($s$)},
                legend style={nodes={scale=1, transform shape}, font=\huge},
            y filter/.code=\pgfmathparse{#1 * 2.30258509299},
            ymode = log,
            log ticks with fixed point,
            ytick = {1, 5, 10, 50, 100, 500},
            y label style={font=\huge},
            x label style={font=\huge},
            xtick = {1, 2, 3, 4, 5},
            legend pos = north west
                ]
                \addplot[color=blue, mark=square*, ultra thin, mark options={solid, fill=blue, scale=1}] coordinates {
(1, 0.08707120591)
(2, 0.387936831)
(3, 0.5641924606)
(4, 0.6891311972)
                };
                \addlegendentry{$|\Proc| = 1; \varphi_6$}
                
                \addplot[color=teal, mark=square*, ultra thin, mark options={solid, fill=teal, scale=1}] coordinates {
(1, -0.1008363585)
(2, 0.2000292666)
(3, 0.3762848962)
(4, 0.5012236328)
                };
                \addlegendentry{$|\Proc| = 1; \varphi_4$}
                
                \addplot[color=black, mark=square*, ultra thin, mark options={solid, fill=black, scale=1}] coordinates {
(1, 0.244029589)
(2, 0.5450525602)
(3, 0.7211508437)
(4, 0.8461527959)
                };
                \addlegendentry{$|\Proc| = 2; \varphi_6$}
                
                \addplot[color=brown, mark=square*, ultra thin, mark options={solid, fill=brown, scale=1}] coordinates {
(1, 0.7911711496)
(2, 1.092194121)
(3, 1.268292404)
(4, 1.393294356)
                };
                \addlegendentry{$|\Proc| = 2; \varphi_4$}
                
                \addplot[color=olive, mark=square*, ultra thin, mark options={solid, fill=olive, scale=1}] coordinates {
(1, 0.612995656)
(2, 0.914025123)
(3, 1.090116382)
(4, 1.215055119)
                };
                \addlegendentry{$|\Proc| = 3; \varphi_6$}
                
                \addplot[color=magenta, mark=square*, ultra thin, mark options={solid, fill=magenta, scale=1}] coordinates {
(1, 1.914571089)
(2, 2.215600556)
(3, 2.391691815)
(4, 2.516630551)
                };
                \addlegendentry{$|\Proc| = 3; \varphi_4$}
                
                \addplot[color=red, mark=square*, ultra thin, mark options={solid, fill=red, scale=1}] coordinates {
(1, 1.044186851)
(2, 1.345216734)
(3, 1.521307806)
(4, 1.646247853)
                };
                \addlegendentry{$|\Proc| = 4; \varphi_6$}
                
                \addplot[color=orange, mark=square*, ultra thin, mark options={solid, fill=orange, scale=1}] coordinates {
(1, 2.587349014)
(2, 2.888378898)
(3, 3.06446997)
(4, 3.189410017)
                };
                \addlegendentry{$|\Proc| = 4; \varphi_4$}

            \end{axis}
        \end{tikzpicture}

%% file: graph_eventRate.tex
        \pgfplotsset{every tick label/.append style={font=\huge}}
    
    \pgfplotsset{every axis/.append style={
                       label style={font=\huge}}}
        \begin{tikzpicture}
            \begin{axis}[
                width=\textwidth,
                grid=both,
                minor tick num=2,
                xlabel={Event Rate ($event/sec$)},
                ylabel={Runtime ($s$)},
                legend style={nodes={scale=1, transform shape}, font=\huge},
            y filter/.code=\pgfmathparse{#1 * 2.30258509299},
            ymode = log,
            log ticks with fixed point,
            ytick = {1, 5, 10, 50, 100, 500},
            y label style={font=\huge},
            x label style={font=\huge},
            xtick = {5, 7, 9, 11, 13, 15},
            legend pos = north west
                ]
                \addplot[color=blue, mark=square*, ultra thin, mark options={solid, fill=blue, scale=1}] coordinates {
(5, 0.6838431062)
(6, 0.7825804826)
(7, 0.8463615581)
(8, 0.9142920276)
(9, 0.9603114569)
(10, 1.087071206)
(13, 1.13073592)
(15, 1.188838394)
                };
                \addlegendentry{$|\Proc| = 1; \varphi_6$}
                
                \addplot[color=teal, mark=square*, ultra thin, mark options={solid, fill=teal, scale=1}] coordinates {
(5, 0.8463371121)
(6, 0.9450744885)
(7, 1.008855564)
(8, 1.076786034)
(9, 1.122805463)
(10, 1.249565212)
(13, 1.293229925)
(15, 1.3513324)
                };
                \addlegendentry{$|\Proc| = 1; \varphi_4$}
                
                \addplot[color=black, mark=square*, ultra thin, mark options={solid, fill=black, scale=1}] coordinates {
(5, 0.8616724561)
(6, 0.9264983332)
(7, 1.042890701)
(8, 1.116935768)
(9, 1.167638604)
(10, 1.244029589)
(13, 1.418723555)
(15, 1.596440743)
                };
                \addlegendentry{$|\Proc| = 2; \varphi_6$}
                
                \addplot[color=brown, mark=square*, ultra thin, mark options={solid, fill=brown, scale=1}] coordinates {
(5, 1.021106568)
(6, 1.085932446)
(7, 1.202324813)
(8, 1.27636988)
(9, 1.327072717)
(10, 1.403463701)
(13, 1.578157667)
(15, 1.755874856)
                };
                \addlegendentry{$|\Proc| = 2; \varphi_4$}
                
                \addplot[color=olive, mark=square*, ultra thin, mark options={solid, fill=olive, scale=1}] coordinates {
(5, 0.8783100994)
(6, 0.9874545689)
(7, 1.130122072)
(8, 1.31196566)
(9, 1.43975224)
(10, 1.612995656)
(13, 1.895164434)
(15, 2.054401358)
                };
                \addlegendentry{$|\Proc| = 3; \varphi_6$}
                
                \addplot[color=magenta, mark=square*, ultra thin, mark options={solid, fill=magenta, scale=1}] coordinates {
(5, 1.146128036)
(6, 1.255272505)
(7, 1.397940009)
(8, 1.579783597)
(9, 1.707570176)
(10, 1.880813592)
(13, 2.162982371)
(15, 2.322219295)
                };
                \addlegendentry{$|\Proc| = 3; \varphi_4$}
                
                \addplot[color=red, mark=square*, ultra thin, mark options={solid, fill=red, scale=1}] coordinates {
(5, 1.084894349)
(6, 1.221114096)
(7, 1.429390038)
(8, 1.69823535)
(9, 1.843567246)
(10, 2.044186851)
(13, 2.381709588)
(15, 2.591470583)
                };
                \addlegendentry{$|\Proc| = 4; \varphi_6$}
                
                \addplot[color=orange, mark=square*, ultra thin, mark options={solid, fill=orange, scale=1}] coordinates {
(5, 1.278753601)
(6, 1.414973348)
(7, 1.62324929)
(8, 1.892094603)
(9, 2.037426498)
(10, 2.238046103)
(13, 2.57556884)
(15, 2.785329835)
                };
                \addlegendentry{$|\Proc| = 4; \varphi_4$}

            \end{axis}
        \end{tikzpicture}

%% file: graph_blockc.tex
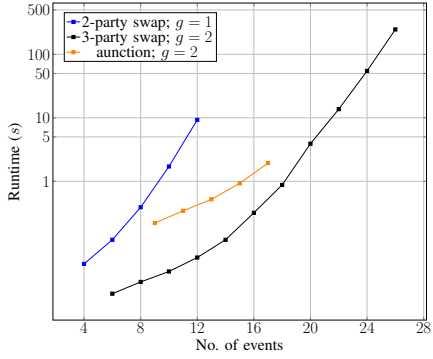
\begin{figure}
    \centering 
    \scalebox{0.3}{
        \pgfplotsset{every tick label/.append style={font=\huge}}
    
    \pgfplotsset{every axis/.append style={
                       label style={font=\huge}}}
        \begin{tikzpicture}
            \begin{axis}[
                width=\textwidth,
                grid=both,
                minor tick num=2,
                xlabel={No. of events},
                ylabel={Runtime ($s$)},
                legend style={nodes={scale=1, transform shape}, font=\huge},
            y filter/.code=\pgfmathparse{#1 * 2.30258509299},
            ymode = log,
            log ticks with fixed point,
            ytick = {1, 5, 10, 50, 100, 500},
            y label style={font=\huge},
            x label style={font=\huge},
            xtick = {1, 4, 8, 12, 16, 20, 24, 28},
            legend pos = north west
                ]
                \addplot[color=blue, mark=square*, ultra thin, mark options={solid, fill=blue, scale=1}] coordinates {
(4, -1.301029996)
(6, -0.920818754)
(8, -0.408935393)
(10, 0.2335037603)
(12, 0.9682026681)
                };
                \addlegendentry{2-party swap; $g = 1$}
                
                \addplot[color=black, mark=square*, ultra thin, mark options={solid, fill=black, scale=1}] coordinates {
(6, -1.769551079)
(8, -1.585026652)
(10, -1.420216403)
(12, -1.200659451)
(14, -0.920818754)
(16, -0.4962093169)
(18, -0.05700040663)
(20, 0.5923988461)
(22, 1.13646689)
(24, 1.738534811)
(26, 2.391749968)
                };
                \addlegendentry{3-party swap; $g = 2$}
                
                \addplot[color=orange, mark=square*, ultra thin, mark options={solid, fill=orange, scale=1}] coordinates {
(9, -0.6575773192)
(11, -0.46470588)
(13, -0.2839966564)
(15, -0.03198428601)
(17, 0.2871296207)
                };
                \addlegendentry{aunction; $g = 2$}

            \end{axis}
        \end{tikzpicture}
}
    \caption{Blockchain Experiments}
    \label{graph:blockc}
     \vspace{-5mm}
\end{figure}

%% file: appendix.tex
\section{Appendix}

Here, in Section~\ref{app:uppaal} we explain how the different \code{UPPAAL} models work and in Section~\ref{app:blockc} we dive into the \MTL specifications we use to verify 3-party swap and the auction protocol.

\subsection{\code{UPPAAL} Models}
\label{app:uppaal}

Below we explain in details how each of the \code{UPPAAL} models work. In respect to our monitoring algorithm, we consider multiple instances of each of the models as different processes. Each event consists of the action that was taken along with the time of occurrence of the event. In addition to this, we assume a unique clock for each instance, synchronized by the presence of a clock synchronization algorithm with a maximum clock skew of $\epsilon$.

\paragraph{The Train-Gate} It models a railway control system which controls access to a bridge for several trains. The bridge can be considered as a shared resource and can be accessed by one train at a time. Each train is identified by a unique \texttt{id} and whenever a new train appears in the system, it sends a \texttt{appr} message along with it's \texttt{id}. The Gate controller has two options: (1) send a \texttt{stop} message and keep the train in waiting state or (2) let the train cross the bridge. Once the train crosses the bridge, it sends a \texttt{leave} message signifying the bridge is free for any other train waiting to cross.
\input{train}
The gate keeps track of the state of the bridge, in other words the gate acts as the controller of the bridge for the trains. If the bridge is currently not being used, the gate immediately offers any train appearing to go ahead, otherwise it sends a \texttt{stop} message. Once the gate is free from a train leaving the bridge, it sends out a \texttt{go} message to any train that had appeared in the mean time and was waiting in the queue.
\input{gate}
\paragraph{The Fischer's Protocol} It is a mutual exclusion protocol designed for $n$ processes. A process always sends in a request to enter the critical section (\texttt{cs}). On receiving the request, a unique \texttt{pid} is generated and the process moves to a \texttt{wait} state. A process can only enter into the critical section when it has the correct \texttt{id}. Upon exiting the critical section, the process resets the \texttt{id} which enables other processes to enter the \texttt{cs}
\input{fischer}
%
\input{gossip}
\paragraph{The Gossiping People} The model consists of $n$ people, each having a private secret they wish to share with each other. Each person can \texttt{Call} another person and after a conversation, both person mutually knows about all their secrets. With respect to our monitoring problem, we make sure that each person generates a new secret that needs to be shared among others infinitely often.

\subsection{Blockchain}
\label{app:blockc}

Below shows the specifications we used to verify the correctness of hedged three-party swap and auction protocols, as shown in \cite{xue2021hedging}.
The structure of the specifications are similar to that of hedged two-party swap protocol. 

\subsubsection{Hedged 3-Party Swap Protocol}

The three-party swap example we implemented can be described as a digraph where there are directed edges between Alice, Bob and Carol. For simplicity, we consider each party transfers 100 assets. Transfer between Alice and Bob is called $ApricotSwap$, meaning Alice proposes to transfer 100 apricot tokens to Bob, transfer between Bob and Carol called $BananaSwap$, meaning Bob proposes to transfer 100 banana tokens to Carol, transfer between Carol and Alice, called $CherrySwap$, meaning Carol proposes to transfer 100 cherry tokens to Alice. Different tokens are managed by different blockchains (Apricot, Banana and Cherry respectively).

We denote the time they reach an agreement of the swap as $startTime$. $\Delta$ is the maximum time for parties to observe the state change of contracts by others and take a step to make changes on contracts. According of the protocol, the execution should follow the following steps:
\begin{itemize}
    \item Step 1. Alice deposits  3 tokens as $escrow\_premium$ in  $ApricotSwap$ before $\Delta$ elapses after $startTime$ .
    \item Step 2. Bob deposits  3 tokens as  $escrow\_premium$ in $BananaSwap$ before $2\Delta$ elapses after $startTime$ .
    \item Step 3. Carol deposits 3 tokens as $escrow\_premium$ in $CherrySwap$ before $3\Delta$ elapses after $startTime$.
    \item Step 4. Alice deposits 3 tokens as $redemption\_premium$ in $CherrySwap$ before $4\Delta$ elapses after $startTime$.
    \item Step 5. Carol deposits 2 tokens as $redemption\_premium$ in $BananaSwap$ before $5\Delta$ elapses after $startTime$ .
    \item Step 6. Bob deposits  1 token as $redemption\_premium$ in $ApricotSwap$ before $6\Delta$ elapses after $startTime$. 
    \item Step 7. Alice escrows 100 ERC20 tokens to $ApricotSwap$  before $7\Delta$ elapses after $startTime$.
     \item Step 8. Bob escrows 100 ERC20 tokens to $BananaSwap$  before $8\Delta$ elapses after $startTime$.
     \item Step 9. Carol escrows 100 ERC20 tokens to $CherrySwap$  before $9\Delta$ elapses after $startTime$.
    \item Step 10. Alice sends the preimage of the hashlock to $CherrySwap$ to redeem Carol's 100 tokens before $10\Delta$ elapses after $startTime$. 
    \item Step 11. Carol sends the preimage of the hashlock to $BananaSwap$  to redeem Bob's 100 tokens before $11\Delta$ elapses after $startTime$. 
     \item Step 12. Bob sends the preimage of the hashlock to $ApricotSwap$  to redeem Alice's 100 tokens before $12\Delta$ elapses after $startTime$. 
\end{itemize} 

If all parties are conforming, the protocol is executed as above. Otherwise, some asset refund and premium redeem events will be triggered to resolve the case where some party deviates. To avoid distraction, we do not provide details here.

\paragraph{Liveness}
Below shows the specification to liveness, if all the steps of the protocol has been taken:
\begin{align*}
&\varphi_{\mathsf{liveness}}=\F_{[0, \Delta)} \texttt{apr.depositEscrowPr(alice)}\\
    & \land  \F_{[0, 2\Delta)} \texttt{ban.depositEscrowPr(bob)}  \\
    & \land \F_{[0, 3\Delta)} \texttt{che.depositEscrowPr(carol)}  \\
    & \land \F_{[0, 4\Delta)} \texttt{che.depositRedemptionPr(alice)}  \\
    & \land \F_{[0, 5\Delta)} \texttt{ban.depositRedemptionPr(carol)}  \\
    & \land \F_{[0, 6\Delta)} \texttt{apr.depositRedemptionPr(bob)} \\
    & \land \F_{[0, 7\Delta)} \texttt{apr.assetEscrowed(alice)}  \\
    & \land \F_{[0, 8\Delta)} \texttt{ban.assetEscrowed(bob)}  \\
    & \land \F_{[0, 9\Delta)} \texttt{che.assetEscrowed(carol)}  \\
    & \land \F_{[0, 10\Delta)} \texttt{che.hashlockUnlocked(alice)}  \\
    & \land \F_{[0, 11\Delta)} \texttt{ban.hashlockUnlocked(carol)}  \\
    & \land \F_{[0, 12\Delta)} \texttt{apr.hashlockUnlocked(bob)} \\
    & \land \F \texttt{assetRedeemed(alice)}  \\
    & \land \F \texttt{assetRedeemed(bob)}  \\
    & \land \F \texttt{assetRedeemed(carol)}  \\
     & \land \F \texttt{EscrowPremiumRefunded(alice)}  \\
      & \land \F \texttt{EscrowPremiumRefunded(bob)}  \\
       & \land \F \texttt{EscrowPremiumRefunded(carol)}  \\
     &\land \F \texttt{RedemptionPremiumRefunded(alice)}\\
      &\land \F \texttt{RedemptionPremiumRefunded(bob)}\\
    &\land \F \texttt{RedemptionPremiumRefunded(carol)}
\end{align*}

\paragraph{Safety}
Below shows the specification to check if an individual party is conforming. If a party is found to be conforming we ensure that there is no negative payoff for the corresponding party.
Specification to check Alice is conforming:
\begin{align*}
    &\varphi_{\mathsf{alice\_conf}} = \F_{[0, \Delta)} \texttt{apr.depositEscrowPr(alice)} \\
    & \land  \big( \F_{[0, 3\Delta)} \texttt{che.depositEscrowPr(carol)} \rightarrow \\
    & \F_{[0, 4\Delta)} \texttt{che.depositRedemptionPr(alice)} \big) \\
    & \land \big( \neg \texttt{che.depositRedemptionPr(alice)} \U \\
    & \texttt{che.depositEscrowPr(carol)} \big) \land \\
    & \big( \F_{[0, 6\Delta)} \texttt{apr.depositRedemptionPr(bob)} \rightarrow \\
    & \F_{[0, 7\Delta)} \texttt{apr.assetEscrowed(alice)} \big)  \\
    & \land \big( \neg \texttt{apr.assetEscrowed(alice)} \U \\
    & \texttt{apr.depositRedemptionPr(bob)} \big)  \\
    & \land \big( \F_{[0, 9\Delta)} \texttt{che.assetEscrowed(carol)} \rightarrow \\
    &  \F_{[0, 10\Delta)} \texttt{che.hashlockUnlocked(alice)} \big)  \\
    & \land \big( \neg \texttt{che.hashlockUnlocked(alice)} \U \\
    & \texttt{che.assetEscrowed(carol)} \big) \land \\
    & \big( \neg \texttt{ban.hashlockUnlocked(carol)} \U \\
    & \texttt{che.hashlockUnlocked(alice)} \big)  \\
    & \land \big( \neg \texttt{apr.hashlockUnlocked(bob)} \U \\
    & \texttt{che.hashlockUnlocked(alice)} \big)
\end{align*}
Specification to check conforming Alice does not have a negative payoff:
\begin{align*}
\varphi_{\mathsf{alice\_safety}} &= \varphi_{\mathsf{alice\_conform}} \rightarrow \\
\big( \sum_{\texttt{TransTo = alice}} \texttt{amount} &\geq \sum_{\texttt{TransFrom = alice}} \texttt{amount} \big)
\end{align*}

\paragraph{Hedged}
Below shows the specification to check that, if a party is conforming and its escrowed asset is refunded, then it gets a premium as compensation.
\begin{align*}
\varphi_{\mathsf{alice\_hedged}}= &\F \big( \varphi_{\mathsf{alice\_conform}}\\
&\land \texttt{apr.assetEscrowed(alice)} \big) \\
&\rightarrow \F \big( \sum_{\texttt{TransTo = alice}} \texttt{amount} \\
&\geq \sum_{\texttt{TransFrom = alice}} \texttt{amount}\\
&+\texttt{apr.redemptionPremium.amount} \big)
\end{align*}

\subsubsection{Auction Protocol}

In the auction example, we consider Alice to be the auctioneer who would like to sell a ticket (worth 100 ERC20 tokens) on the ticket (\texttt{tckt}) blockchain, and Bob and Carol bid on the \texttt{coin} blockchain and the winner should get the ticket and pay for the auctioneer what they bid, and the loser will get refunded. We denote the time that they reach an agreement of the auction as $startTime$. $\Delta$ is the maximum time for parties to observe the state change of contracts by 
others and take a step to make changes on contracts. Let $TicketAuction$ be a contract managing the ``ticket" on the ticket blockchain, and $CoinAuction$ be a contract managing the bids on the coin blockchain. The protocol is briefed as follows.

\begin{itemize}
    \item Setup.  Alice generates two hashes $h(s_b)$ and $h(s_c)$. $h(s_b)$ is assigned to Bob and  $h(s_c)$ is assigned to Carol. If Bob is the winner, then Alice releases $s_b$. If Carol is the winner, then Alice releases $s_c$.  If both $s_b$ and $s_c$ are released in $TicketAuction$, then the ticket is refunded. If both $s_b$ and $s_c$ are released in  $CoinAuction$ , then all coins are refunded. In addition, Alice escrows her ticket as 100 ERC20 tokens in $TicketAuction$ and deposits 2 tokens as premiums in $CoinAuction$.
    \item Step 1 (Bidding).   Bob and Carol bids before $\Delta$ elapses after $startTime$.
    \item Step 2 (Declaration). Alice sends the winner's secret to both chains to declare a winner before $2\Delta$ elapses after $startTime$.
    \item Step 3 (Challenge). Bob and Carol challenges if they see two secrets or one secret missing, i.e. Alice cheats,  before $4\Delta$ elapses after $startTime$. They challenge by forwarding the secret released by Alice using a path signature scheme \cite{herlihy2018atomic}.
    \item Step 4 (Settle). After  $4\Delta$ elapses after $startTime$, on the $CoinAuction$,  if only the  hashlock corresponding to the actual winner is unlocked, then the winner's bid goes to Alice. Otherwise, the winner's bid is refunded. Loser's bid is always refunded. If the winner's bid is refunded, all bidders including the loser gets 1 token as premium to compensate them. On the $TicketAuction$, if only one secret is released, then the ticket is transferred to the corresponding party who is assigned the hash of the secret. Otherwise, the ticket is refunded.
\end{itemize}

\paragraph{Liveness}
Below shows the specification to check that, if all parties are conforming, the winner (Bob) gets the ticket and the auctioneer gets the winner's bid. 
\begin{align*}
    \varphi_{\mathsf{liveness}} &= \F_{[0, \Delta)} \texttt{coin.bid(bob)} \\
    &\land   \F_{[0, 2\Delta)} \texttt{coin.declaration(alice, $s_b$)}  \\
    &\land  \F_{[0, 2\Delta)} \texttt{tckt.declaration(alice, $s_b$)}  \\
    &\land  \F_{(4\Delta,\infty)} \texttt{coin.redeemBid(any)} \\
    & \land \F_{(4\Delta,\infty)} \texttt{coin.refundPremium(any)} \\
    &\land  \big(\texttt{coin.bid(carol)} \rightarrow \\
    &\F_{[0, \Delta)} \texttt{coin.refundBid(any)}\big)  \\
    &\land  \texttt{tckt.redeemTicket(any)}  \\
    & \land \neg \texttt{coin.challenge(any)} \\
    & \land \neg  \texttt{tckt.challenge(any)}
\end{align*}
\paragraph{Safety}
Below shows the specification to check that, if a party is conforming, this party does not end up worse off. Take Bob (the winner) for example.

Specification to define Bob is conforming:
\begin{align*}
    \varphi_{\mathsf{bob\_conform}} &= \F_{[0, \Delta)} \texttt{coin.bid(bob)} \\
    & \land \Big( \big( \texttt{coin.declaration(alice, $s_c$)} \lor \\
    & \texttt{coin.challenge(carol, $s_c$)} \big) \rightarrow \\
    & \land \big( \texttt{tckt.declaration(alice, $s_c$)} \lor \\
    & \texttt{tckt.challenge(carol, $s_c$)} \lor \\
    & \texttt{tckt.challenge(bob, $s_c$)} \big) \Big) \\
    & \land \Big( \big( \texttt{coin.declaration(alice, $s_b$)} \lor \\
    & \texttt{coin.challenge(carol, $s_b$)} \big) \rightarrow \\
    & \land \big( \texttt{tckt.declaration(alice, $s_b$)} \lor \\
    & \texttt{tckt.challenge(carol, $s_b$)} \lor \\
    & \texttt{tckt.challenge(bob, $s_b$)} \big) \Big) \\
    & \land \Big( \big( \texttt{tckt.declaration(alice, $s_c$)} \lor \\
    & \texttt{tckt.challenge(carol, $s_c$)} \big) \rightarrow \\
    & \land \big( \texttt{coin.declaration(alice, $s_c$)} \lor \\
    & \texttt{coin.challenge(carol, $s_c$)} \lor \\
    & \texttt{coin.challenge(bob, $s_c$)} \big) \Big) \\
    & \land \Big( \big( \texttt{tckt.declaration(alice, $s_b$)} \lor \\
    & \texttt{tckt.challenge(carol, $s_b$)} \big) \rightarrow \\
    & \land \big( \texttt{coin.declaration(alice, $s_b$)} \lor \\
    & \texttt{coin.challenge(carol, $s_b$)} \lor \\
    & \texttt{coin.challenge(bob, $s_b$)} \big) \Big)
\end{align*}
Specification to define Bob does not end up worse off:
\begin{align*}
    \varphi_{\mathsf{bob\_safety}} &= \varphi_{\mathsf{bob\_conform}} \rightarrow \\
    & \F \Big( \big( \texttt{coin.refundBid(any)} \\
    & \land \texttt{coin.redeemPremium(any)}\big) \lor \\
    & \texttt{tckt.redeemTicket(any)} \Big)
\end{align*}
\paragraph{Hedged}
Below shows the specification to check that, if a party is conforming and its escrowed asset is refunded, then it gets a premium as compensation.
\begin{align*}
    \varphi_{\mathsf{bob\_hedged}} &= \G \Big(\varphi_{\mathsf{bob\_conforming}} \\
    & \land \big( \texttt{tckt.refundTicket(alice)}  \lor \\
    & \texttt{tckt.redeemTicket(carol)} \big) \Big) \rightarrow \\
    & \F \big( \texttt{coin.refundBid(any)} \\
    & \land \texttt{coin.redeemPremium(any)} \big)
\end{align*}

%% file: train.tex
\begin{figure}[H]
    \tikzset{
        ->,
        >=stealth,
        node distance=75,
        every state/.style={thick,
        fill=gray!10},
        initial text=$ $
    }
    \tikzstyle{every node}=[font=\small]
    \centering
    \scalebox{.7}{
    \begin{tikzpicture}
        
        \node[state, minimum size=1cm, initial] (1) {$\texttt{Safe}$};
        \node[state, minimum size=1cm, below of=1] (2) {$\texttt{Appr}$};
        \node[state, minimum size=1cm, below right of=2] (3) {$\texttt{Stop}$};
        \node[state, minimum size=1cm, above right of=3] (4) {$\texttt{Start}$};
        \node[state, minimum size=1cm, above of=4] (5) {$\texttt{Cross}$};
        
        \draw (1) edge node[left]{$\texttt{appr[id]}$} (2);
        \draw (2) edge node[left]{$\texttt{stop[id]}$} (3);
        \draw (3) edge node[right]{$\texttt{go[id]}$} (4);
        \draw (4) edge node[left]{} (5);
        \draw (5) edge node[above]{$\texttt{leave[id]}$} (1);
        \draw (2) edge node[left]{} (5);
        
    \end{tikzpicture}
    }
    \caption{Train model}
    \label{fig:train}
\end{figure}
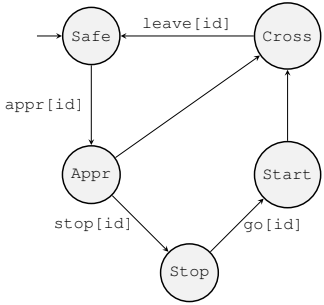

%% file: gate.tex
\begin{figure}[H]
    \tikzset{
        ->,
        >=stealth,
        node distance=75,
        every state/.style={thick,
        fill=gray!10},
        initial text=$ $
    }
    \tikzstyle{every node}=[font=\small]
    \centering
    \scalebox{.7}{
    \begin{tikzpicture}
        
        \node[state, minimum size=1cm, initial] (1) {$\texttt{Free}$};
        \node[state, minimum size=1cm, below of=1] (2) {$\texttt{Occ}$};
        \node[state, minimum size=1cm, below of=2] (3) {};
        
        \draw (1) edge [bend right=90] node[left]{$\texttt{go[front()]}$} (2);
        \draw (1) edge node[above]{$\texttt{appr[e]}$} (2);
        \draw (2) edge [bend right=90] node[right]{$\texttt{leave[id]}$} (1);
        \draw (2) edge [bend right] node[left]{$\texttt{appr[e]}$} (3);
        \draw (3) edge [bend right] node[right]{$\texttt{stop[tail()]}$} (2);
        
    \end{tikzpicture}
    }
    \caption{Gate model}
    \label{fig:gate}
\end{figure}
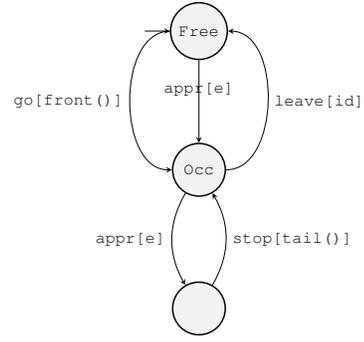

%% file: fischer.tex
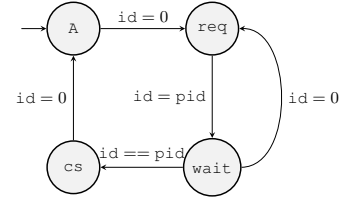
\begin{figure}[H]
    \tikzset{
        ->,
        >=stealth,
        node distance=75,
        every state/.style={thick,
        fill=gray!10},
        initial text=$ $
    }
    \tikzstyle{every node}=[font=\small]
    \centering
    \scalebox{.7}{
    \begin{tikzpicture}
        
        \node[state, minimum size=1cm, initial] (1) {$\texttt{A}$};
        \node[state, minimum size=1cm, below of=1] (2) {$\texttt{cs}$};
        \node[state, minimum size=1cm, right of=2] (3) {$\texttt{wait}$};
        \node[state, minimum size=1cm, above of=3] (4) {$\texttt{req}$};
        
        \draw (2) edge node[left]{$\texttt{id} = 0$} (1);
        \draw (3) edge node[above]{$\texttt{id} == \texttt{pid}$} (2);
        \draw (3) edge [bend right=90] node[right]{$\texttt{id} = 0$} (4);
        \draw (4) edge node[left]{$\texttt{id} = \texttt{pid}$} (3);
        \draw (1) edge node[above]{$\texttt{id} = 0$} (4);
        
    \end{tikzpicture}
    }
    \caption{Fischer model}
    \label{fig:fischer}
\end{figure}

%% file: gossip.tex
\begin{figure}[H]
    \tikzset{
        ->,
        >=stealth,
        node distance=75,
        every state/.style={thick,
        fill=gray!10},
        initial text=$ $
    }
    \tikzstyle{every node}=[font=\small]
    \centering
    \scalebox{.7}{
    \begin{tikzpicture}
        
        \node[state, minimum size=1cm, initial] (1) {$\texttt{Start}$};
        \node[state, minimum size=1cm, below of=1] (2) {$\texttt{Call}$};
        \node[state, minimum size=1cm, below of=2] (3) {$\texttt{Listen}$};
        
        \draw (1) edge node[left]{$\texttt{start()}$} (2);
        \draw (2) edge node[left]{$\texttt{talk()}$} (3);
        \draw (3) edge [bend right=90] node[right]{$\texttt{listen()}$} (2);
        \draw (2) [loop left] edge node[left]{$\texttt{exchange()}$} (2);

    \end{tikzpicture}
    }
    \caption{Gossiping people model}
    \label{fig:gossip}
\end{figure}
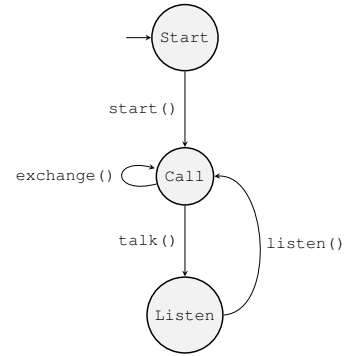